\documentclass[letterpaper,times]{IONconf-v2}
\usepackage[]{graphicx}    
\newcommand{\ignore}[1]{}  

\usepackage{amsmath,amssymb,amsfonts}
\usepackage{bm}
\usepackage{algorithm2e}
\usepackage{float}
\usepackage[labelformat=simple]{subcaption}

\usepackage{mathtools}
\usepackage{siunitx}
\usepackage{multirow}
\usepackage{wasysym}
\usepackage{csquotes}

\usepackage[hidelinks]{hyperref}

\usepackage{amsthm}
\usepackage[english]{babel}

\newcolumntype{I}{!{\vrule width 1.5pt}}

\newcommand{\newtext}[1]{\textcolor{black}{#1}}

\articletype{Regular papers}%

\received{xxxxx}
\revised{xxxxx}
\accepted{xxxxx}
\doi{10.1109/xxx.xx.xxxx}
\journalname{NAVIGATION}
\journalvolume{xxxxx}
\journalnumber{xxxxx}

\title{\newtext{Trade-off Analysis for Lunar Augmented Navigation Service (LANS) Constellation Design}}

\author[1]{Keidai Iiyama}

\author[1]{Grace Gao}

\authormark{IIYAMA \textsc{et al}}

\address[1]{Department of Aeronautics and Astronautics, Stanford University, California, United States}

\corres{Grace Gao, Department of Aeronautics and Astronautics, Stanford University, California, United States \\ \email{gracegao@stanford.edu}}

\begin{document}

\abstract[Abstract]{
The establishment of a sustainable human presence on the Moon demands robust positioning, navigation, and timing (PNT) services capable of supporting both surface and orbital operations. 
This paper presents a comprehensive trade-off analysis of lunar frozen-orbit constellations for the Lunar Augmented Navigation Service (LANS), focusing on how the number of satellites and orbital parameters influence coverage, position dilution of precision (PDOP), orbit determination accuracy, receiver noise, and orbit insertion cost.
Three Walker-constellation families based on frozen elliptical and circular orbits are examined to characterize their relative advantages across different semi-major axes and inclinations.
Results show that larger semi-major axes enhance both polar and global coverage, though the optimal inclination depends on the constellation type and target service region. 
The south elliptical lunar frozen orbit (ELFO) Walker constellation provides superior performance for polar coverage and PDOP, whereas the circular lunar frozen orbit (CLFO) Walker configuration achieves the best global uniformity. 
Orbit determination errors and receiver noise both increase with larger semi-major axes and higher inclinations, reflecting weaker geometric observability and reduced received signal power at apolune for eccentric orbits. 
Orbit insertion analysis reveals clear trade-offs among transfer duration, characteristic energy ($C_3$) at trans-lunar injection, and insertion $\Delta V$: shorter transfers require higher insertion $\Delta V$, while low-energy transfers achieve smaller $\Delta V$ at the cost of months-long durations and higher $C_3$.
These findings provide a systematic framework for designing LANS constellations for both regional and global coverage.

}


\maketitle

\section{Introduction}
The prospect of a sustained human presence on the Moon has accelerated in recent years, with more than 40 missions planned between 2024 and 2030~\citep{PlanetarySocietyMoon2024}. 
To support these activities, reliable positioning, navigation, and timing (PNT) services are essential for both surface and orbital operations. 
NASA and its international partners are therefore developing LunaNet~\citep{Israel2020}, a “network of networks’’ that integrates communication, navigation, detection, and science services throughout the lunar environment. 
A cornerstone of this framework is the Lunar Augmented Navigation Service (LANS)~\citep{LNISv4}, envisioned as the lunar counterpart to terrestrial GNSS. 
Several agencies have proposed satellite constellations for LANS, including NASA’s Lunar Communications Relay and Navigation System (LCRNS)~\citep{gramling2024}, ESA’s Moonlight Lunar Communications and Navigation System (LCNS)~\citep{traveset2024}, and JAXA’s Lunar Navigation Satellite System (LNSS)~\citep{murata2024}.

Designing a lunar navigation constellation entails two tightly coupled aspects: 
(i)~the space segment design, which defines the orbits and satellite configuration, and 
(ii)~the user positioning performance evaluation, which quantifies service quality at the user end. 
From the perspective of the space segment, the primary design choice lies in selecting the orbit family and arranging satellites within it. 
Candidate orbits include elliptic lunar frozen orbits (ELFOs)~\citep{ely2005stable}, circular lunar frozen orbits (CLFOs)~\citep{kang2025itm}, and halo orbits~\citep{Circi2014}, typically organized using constellation schemes such as Walker~\citep{Walker1970} or Flower~\citep{Mortari2004Flower} constellations.
Mission planners must also consider orbit insertion and maintenance costs, which vary significantly across orbit families and parameters.

User positioning performance is commonly evaluated using metrics such as coverage, geometry (e.g., geometric and position dilution of precision, GDOP and PDOP), user range error (URE), and robustness to satellite failures. 
Coverage quantifies the fraction of time a user within a service region can access at least four satellites for 3D positioning and timing. 
DOP metrics measure how satellite geometry amplifies range measurement errors, while URE aggregates uncertainties arising from orbit determination (OD), clock stability, and receiver noise—the latter strongly influenced by received signal power. 
Satellite failures, inevitable over multi-year mission lifetimes, further degrade service availability and must therefore be incorporated into constellation design.

Table~\ref{tab:comparison} summarizes how existing studies have approached LANS constellation design and user performance evaluation. 
Prior works have proposed candidate constellations for global lunar navigation~\citep{Circi2014, Pereira_Navigation_2021, ARCIAGIL_2023, kang2025itm}, and for regional coverage around the south pole~\citep{BhamidipatiConstellation2023, Ceresoli2025}.
Some of the works developed multi-objective optimization frameworks~\citep{Pereira_Navigation_2021, ARCIAGIL_2023,Ceresoli2025} to identify optimal constellation parameters that balance competing performance metrics.
However, these studies exhibit four main limitations. 
First, most evaluate performance solely based on geometric metrics (GDOP/PDOP), without explicitly modeling OD errors, clock stability, or receiver noise—all of which critically influence URE and depend on orbital characteristics. 
Second, except for \citet{kang2025itm}, prior works focus exclusively on either global or polar coverage, without addressing both simultaneously. 
Third, although several studies consider station-keeping costs, they neglect orbit insertion $\Delta V$, which can dominate total propellant expenditure, especially for stable frozen orbits. 
Finally, the use of evolutionary multi-objective optimization often results in black-box outcomes that obscure physical insight into the trade-offs between design parameters.

\begin{table}[htb]
\caption{Comparison of Constellation Design and Evaluation Approaches in Related Works on LANS Constellation Design (SP = South Pole, LOI = Lunar Orbit Insertion, SK = Station-Keeping, OD=Orbit Detemination).}
\label{tab:comparison}
\begin{tblr}{
  colspec={X[2.0cm, l] | X[c]X[c] X[0.5cm, c]X[1.2cm, c] | X[0.5cm, c]X[0.5cm, c]X[c]X[2.5cm, c]X[c] | X[1.8cm, c]},
  width=\textwidth,
  row{odd} = {white, font=\small},
  row{even} = {bg=black!10, font=\small},
  row{1, 2} = {bg=black!20, font=\bfseries\small},
  hline{Z} = {1pt, solid, black!60},
  rowsep=3pt
}
\SetCell[r=2]{c}\textbf{Paper} & 
\SetCell[c=4]{c}\textbf{Space Segment Design} & & & &
\SetCell[c=5]{c}\textbf{User Performance Evaluation} & & & & &
\SetCell[r=2]{c}\textbf{Optimization} \\
 & Orbits & Intended Coverage & LOI $\Delta$V & SK $\Delta$V & DOP & OD & Time Sync & Receiver Noise & Satellite Failure & \\
\citet{Circi2014} & Halo & Global &  &  & \checkmark &  &  & &  & \\
\citet{Pereira_Navigation_2021} & Two-body & Global &  & \checkmark & \checkmark & &  & $\triangle$ (link-budget analysis) & \checkmark & \checkmark \\
\citet{ARCIAGIL_2023} & Circular & Global &  & \checkmark & \checkmark &  &  & & & \checkmark  \\
\citet{BhamidipatiConstellation2023} & ELFO & SP &  & - (frozen orbits) & \checkmark &  &  \checkmark (GPS) & $\triangle$ (fixed value) & \checkmark &  \\
\citet{kang2025itm} & Frozen & SP + Global &  & - (frozen orbits) & \checkmark &  &  &  &  & \\
\citet{Ceresoli2025} & Frozen & SP &  & \checkmark & \checkmark &  &  &  &  & \checkmark \\
\textbf{This work} & Frozen & SP + Global & \checkmark & - (frozen orbits) & \checkmark & \checkmark & - (Two-way) & \checkmark & \checkmark &  \\
\end{tblr}
\end{table}

This paper addresses these gaps through three main contributions. 
Building on our preliminary analysis presented at the 2025 ION GNSS+ Conference~\citep{iiyama2025constellation}, this work extends our previous simulations and adds new analyses, including constellation robustness analysis (Section~\ref{sec:fault_robustness}) and orbit insertion cost modeling (Section~\ref{sec:insertion_cost}). 
\begin{enumerate}
    \item We perform a comprehensive coverage and PDOP analysis for both the lunar south-pole region and global surface, comparing three frozen orbit constellation types. We identify the minimum number of satellites required to maintain 4-fold coverage and PDOP~$\leq$~6 and PDOP $\leq$~3, and analyze the impact of single-satellite loss on system performance.
    \item We investigate how frozen orbit parameters—semi-major axis, inclination, and right ascension of the ascending node (RAAN)—affect orbit determination accuracy and receiver noise, both key contributors to URE. 
    \item We evaluate the orbit insertion $\Delta V$ for each frozen orbit type using both fast direct transfers and low-energy transfers that exploit the Earth–Sun–Moon dynamics, characterizing the relationships between transfer energy ($C_3$), duration, and final orbit geometry.
\end{enumerate}

Optimization of constellation architectures is beyond the scope of this paper and will be addressed in future work.

The remainder of this paper is organized as follows. 
Section~\ref{sec:coverage_dop} analyzes coverage and PDOP performance for various frozen orbit configurations. 
Section~\ref{sec:od_analysis} evaluates orbit determination errors based on a ground-based ODTS approach. 
Section~\ref{sec:receiver_noise} assesses receiver noise through link-budget simulations. 
Section~\ref{sec:insertion_cost} quantifies orbit insertion $\Delta V$ and transfer duration across frozen orbit geometries. 
Finally, Section~\ref{sec:conclusion} summarizes the trade-offs among performance metrics and discusses implications for future constellation optimization.
\section{Coverage and PDOP Analysis}
\label{sec:coverage_dop}
We first perform grid-search analyses to determine the number of satellites
required to achieve four-satellite (4-fold) coverage globally and in the
South Pole region. We also assess how the semi-major axis and inclination affect
position dilution of precision (PDOP).

\subsection{Frozen Orbit Walker Constellation Families}
\begin{figure}[ht!]
  \centering
  \begin{minipage}{0.33\textwidth}
    \centering
    \includegraphics[width=\textwidth,trim={20mm 0mm 20mm 10mm}]
      {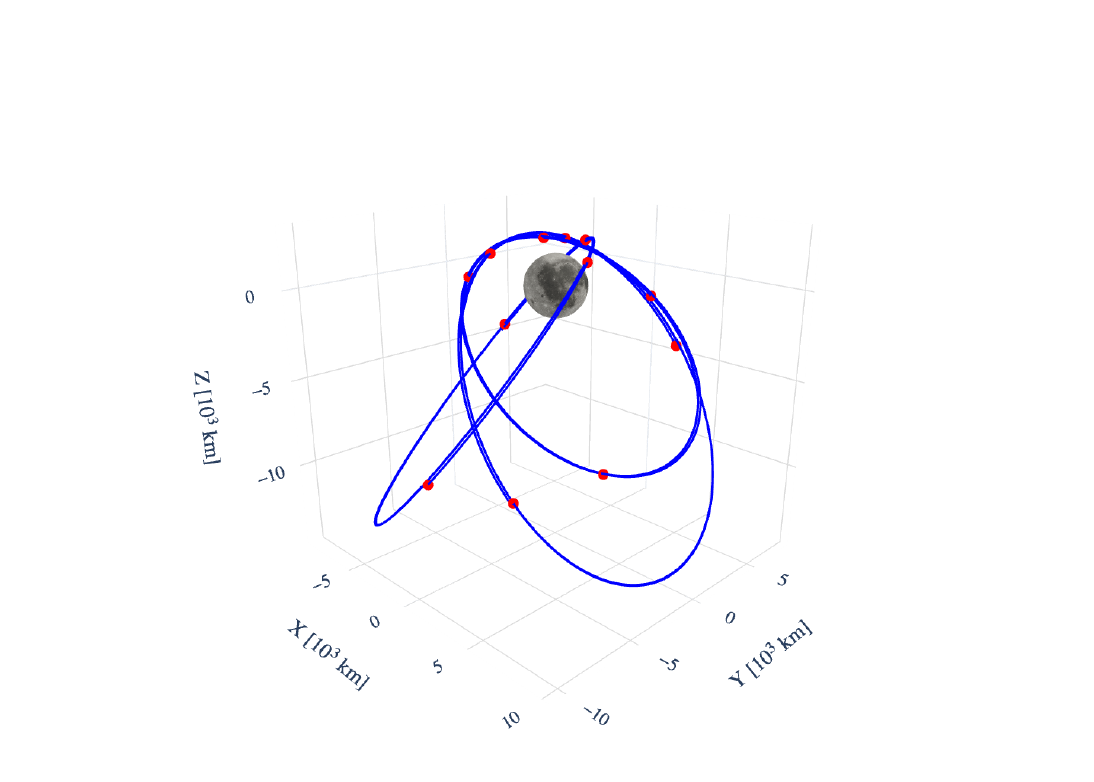}
    \subcaption{S-ELFO Walker $55^{\circ}: 12/3/1$}
  \end{minipage}\hfill
  \begin{minipage}{0.33\textwidth}
    \centering
    \includegraphics[width=\textwidth,trim={20mm 0mm 20mm 10mm}]
      {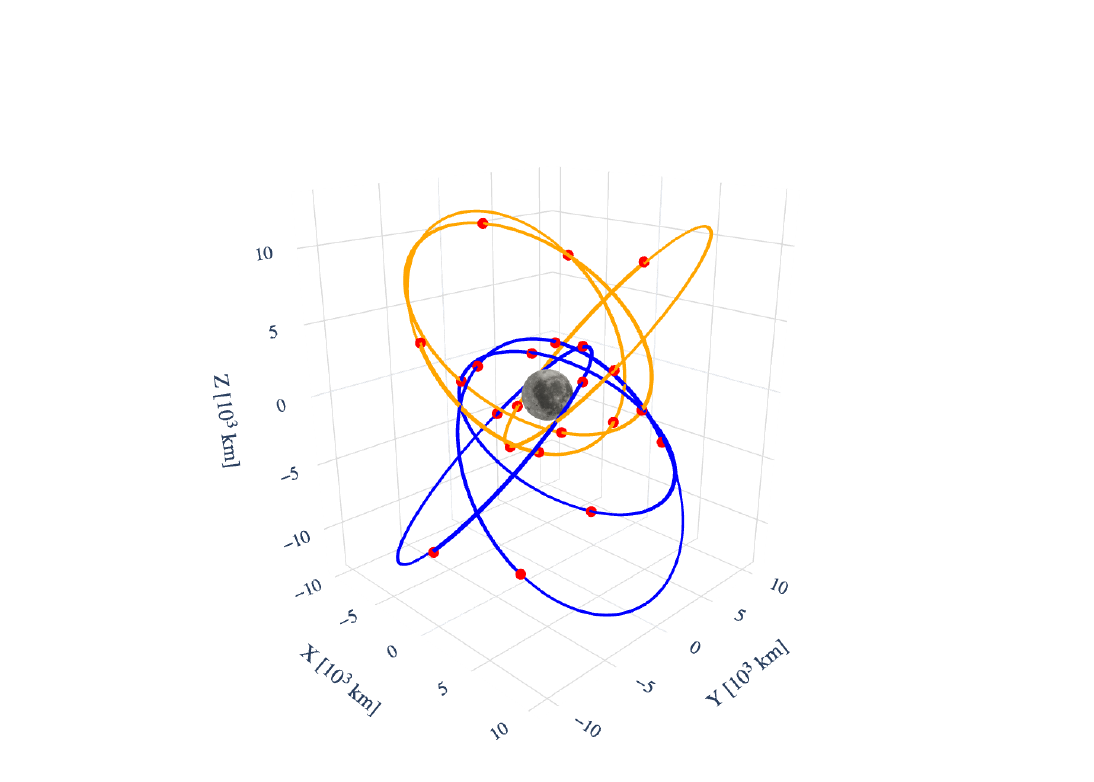}
    \subcaption{NS-ELFO Walker $50^{\circ}: 24/6/1$}
  \end{minipage}\hfill
  \begin{minipage}{0.33\textwidth}
    \centering
    \includegraphics[width=\textwidth,trim={20mm 0mm 20mm 10mm}]
      {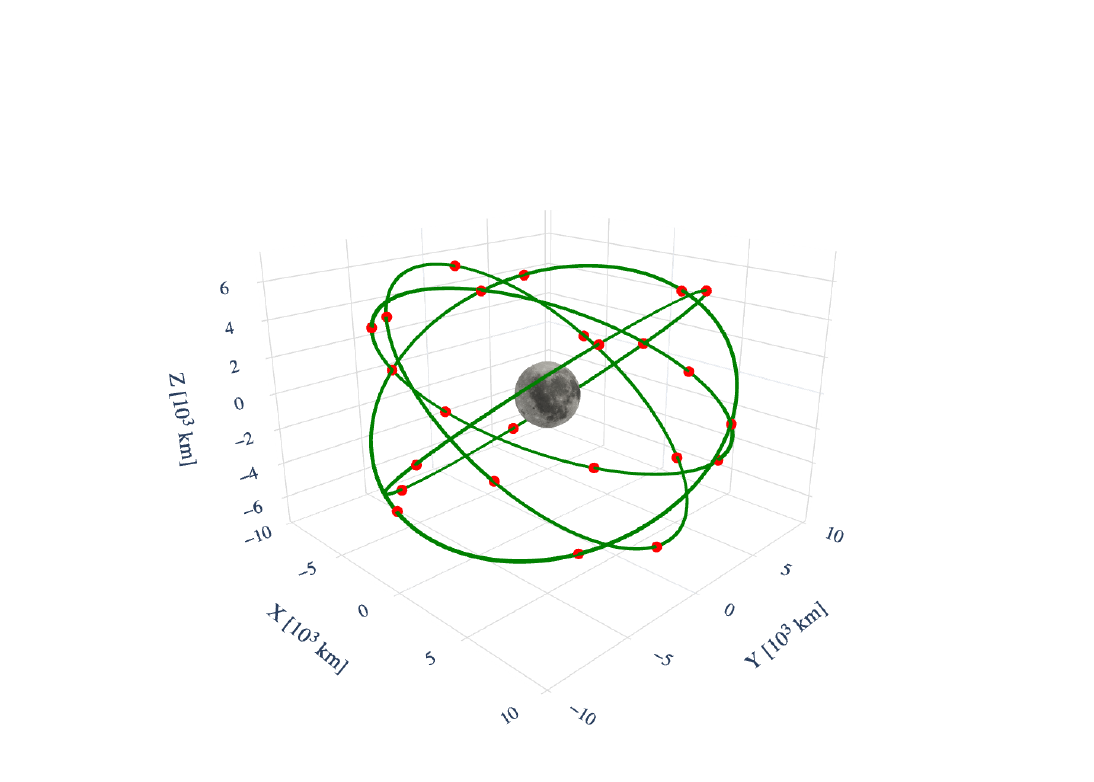}
    \subcaption{CLFO Walker $39.2^{\circ}: 24/4/1$}
  \end{minipage}
  \caption{Examples of the three constellation types used in the grid search.
  The semi-major axis is 9751 km (24-hour orbit) in all cases. Red dots show
  satellite positions at the initial epoch; blue ($\omega = 90^\circ$) and
  orange ($\omega = 270^\circ$) lines show the orbits in an inertial frame
  aligned with the lunar principal axes.}
  \label{fig:constellation_examples}
\end{figure}

We consider three Walker-like phased constellations using frozen orbits:
(i) the \emph{South-ELFO Walker (S-ELFO)}, (ii) the \emph{North+South-ELFO
Walker (NS-ELFO)}, and (iii) the \emph{Circular Lunar Frozen Orbit (CLFO)
Walker}. Constellation configurations are denoted by
$i^{op}\!:\, N_s/N_p/f$, where $i^{op}$ is the inclination in the Earth
Orbit Plane (OP) frame~\citep{ely2005stable}, $N_s$ is the total number of
satellites, $N_p$ is the number of planes (number of satellites per plane is $N_{spp} = N_s / N_p$), and $f$ is the relative phasing between the satellites in adjacent planes.
The initial mean anomaly $M$ of the $i$th satellite in plane $p$ is given by
\begin{align}
  M_{i,p} = 2\pi \frac{i-1}{N_{spp}} + 2\pi \frac{p\,f}{N_s},
\end{align}
We vary the semi-major axis $a$ and inclination $i^{op}$ to identify configurations that perform well in coverage and PDOP. 
The eccentricity $e$ is determined from the frozen-orbit condition
\begin{equation}
  e^2 + \frac{5}{3}\cos^2(i^{op}) = 1.
  \label{eq:frozen_condition}
\end{equation}
For S-ELFO, we set the argument of periapsis $\omega = 90^\circ$ so that apolune remains in the southern hemisphere. 
For NS-ELFO, half of the satellites use $\omega = 90^\circ$ and the other half $\omega = 270^\circ$ to improve equatorial and northern-hemisphere coverage. 
Note that the phasing of the $\omega=90^\circ$ subset in NS-ELFO with $2N_s/N_p/f$ is equivalent to the S-ELFO $N_s/N_p/f$ case, yielding symmetry about the equatorial plane; we therefore only consider even $N_p$ for NS-ELFO. 
For CLFO, we set $e=0$, which gives $i^{op} = 39.23^\circ$ from \eqref{eq:frozen_condition}, and vary only $a$. Examples appear in Figure~\ref{fig:constellation_examples}.

\subsection{Coverage Evaluation}

\begin{figure}[ht!]
  \centering
  \includegraphics[width=0.5\columnwidth]
    {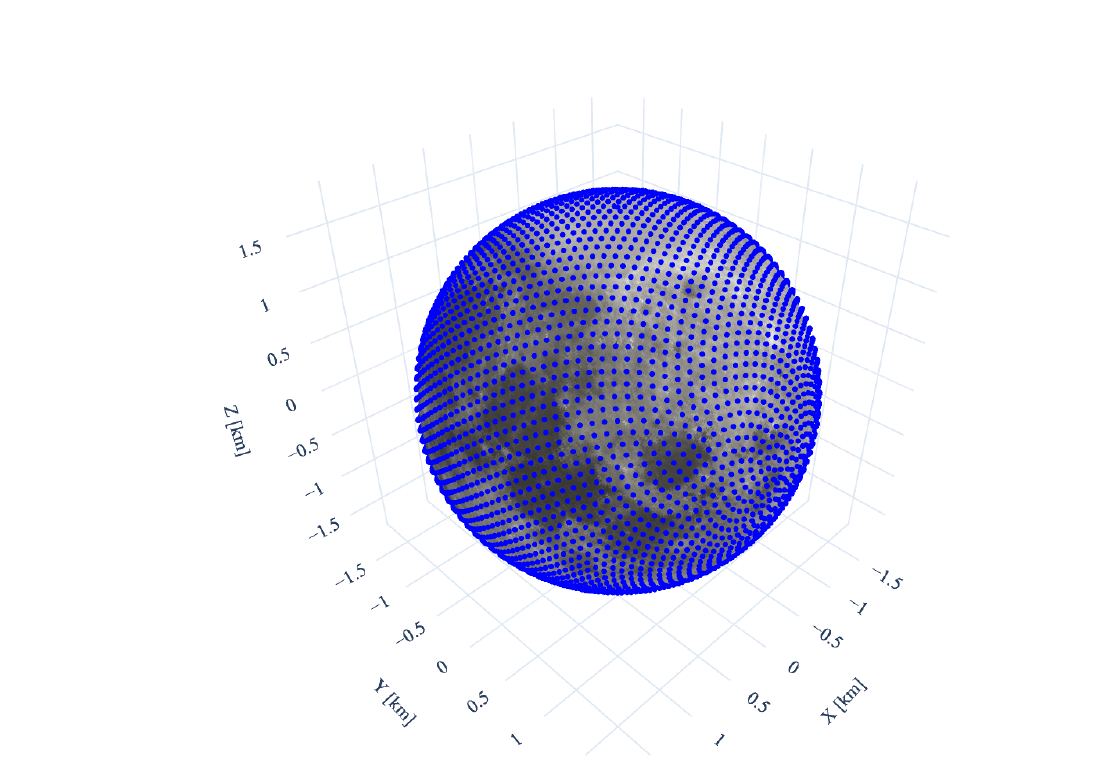}
  \caption{Five thousand user points on the lunar surface sampled with a
  Fibonacci-sphere algorithm.}
  \label{fig:user_points}
\end{figure}

We evaluate surface coverage using 5000 uniformly distributed user points
generated by a Fibonacci-sphere sampler, as shown in Figure~\ref{fig:user_points}. The
simulation spans one orbital period with 1-minute sampling. A satellite is
considered visible if its elevation angle exceeds $5^\circ$. Because both
space and time are discretized, we define ``full 4-fold coverage'' as having
at least four satellites visible for $\geq 99\%$ of the time across
$\geq 99\%$ of user points.

For each configuration, we sweep $a \in [4000,16000]$ km in 1000 km steps and
$i^{op} \in [40^\circ,65^\circ]$ in $2.5^\circ$ steps. 
The upper bound of the semi-major axis was decided so that the apolune radius at its largest fits well within the Hill sphere of the Moon $r_H$, which is
\begin{equation}
    r_H = a_{EM} \left(\frac{\mu_{M}}{3 \mu_{E}}\right)^{1/3} = 61524 \ \text{km}
\end{equation}
where $a_{EM}$ = 38440 km is the Earth-Moon distance, and $\mu_M = 4902 \ \text{km}^3 \text{s}^{-2}, \mu_E =398600 \ \text{km}^3 \text{s}^{-2}$ are the gravitational constants of Earth and Moon, respectively.
Combinations of semi-major axis and inclination (= eccentricity) where perilune falls below the surface are infeasible and shown as white regions in
the plots.

\subsection{South ELFO Walker Constellation}
\begin{figure}[ht!]
  \centering
  \includegraphics[width=\textwidth]
    {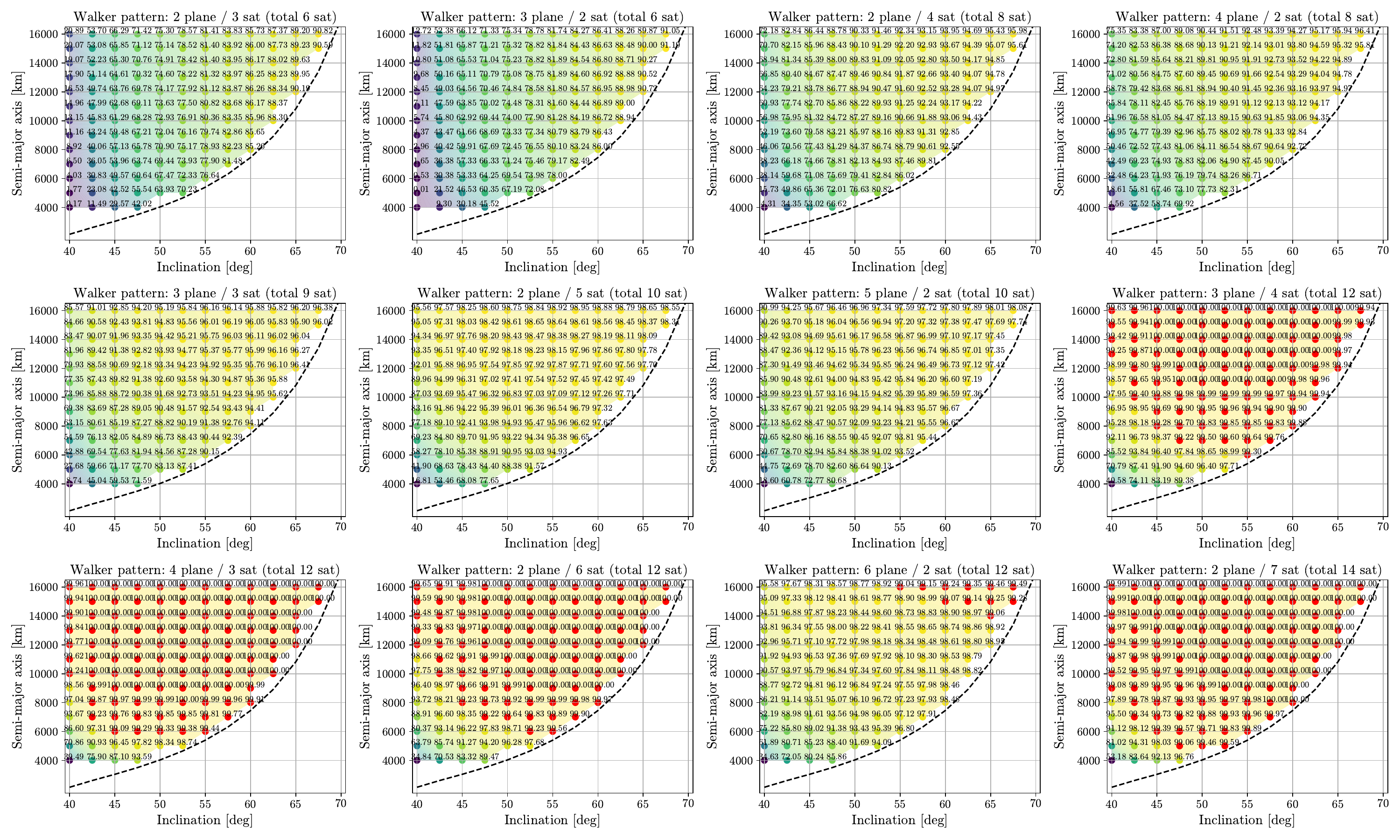}
  \caption{South Pole coverage for S-ELFO as a function of $a$ and $i^{op}$,
  across different $(N_p,N_{spp})$. The color shows the percentage of time and
  user points in the South Pole region (latitude $< \text{S}70^\circ$) with
  4-fold coverage. White regions are infeasible.}
  \label{fig:pole_coverage}
\end{figure}

Figure~\ref{fig:pole_coverage} shows South Pole coverage for S-ELFO over a varying number of planes and satellites per plane. 
Coverage increases with both $a$ and $i^{op}$ across configurations. 
With $a \leq 16{,}000$ km, the minimum to achieve $\geq 99\%$ 4-fold coverage in the South Pole region is 12 satellites, achievable with $(N_p,N_{spp}) \in \{(3,4),(4,3),(2,6)\}$. 
The minimum $a$ to reach $\geq 99\%$ 4-fold coverage at the pole lies between 5000 and
6000 km for all cases. 
If fewer satellites are available, $a=15{,}000$ km and $i^{op}=67.5^\circ$ can still yield $\sim$90\% coverage with 6 satellites and $\sim$95\% with 8 satellites.

\begin{figure}[ht!]
  \centering
  \includegraphics[width=\textwidth]
    {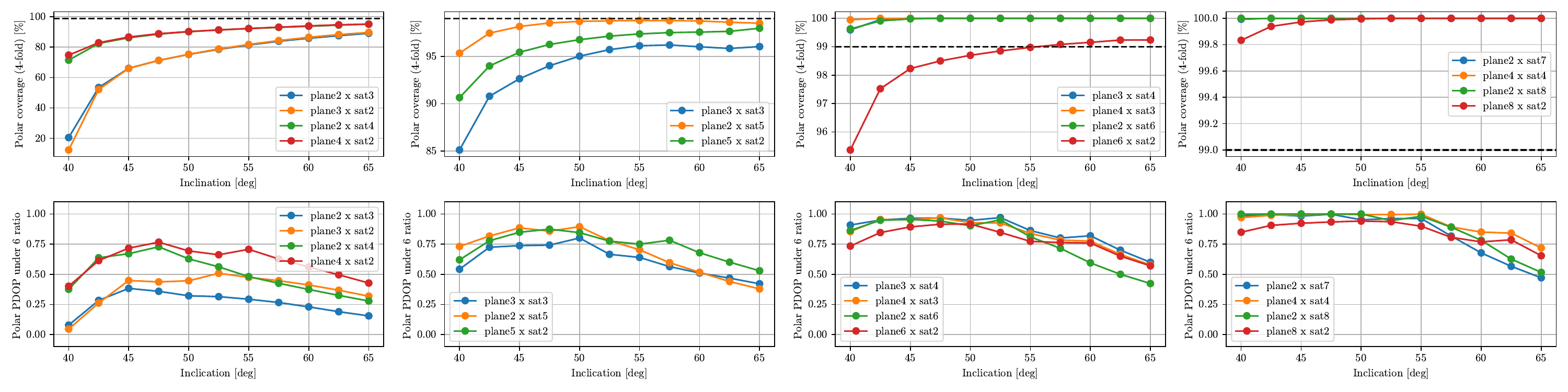}
  \caption{Coverage (top) and fraction of time with PDOP $\leq 6$ in the South
  Pole region (bottom) for S-ELFO at $a=15478$ km (48 hr orbit). Columns correspond to the total
  satellites $\{6,8\}$, $\{9,10,12\}$, and $\{14,16\}$.}
  \label{fig:pole_pdop}
\end{figure}

Figure~\ref{fig:pole_pdop} shows PDOP for S-ELFO as a function of $i^{op}$
with fixed $a=15478$ km (48-hour orbit). While coverage generally improves
with inclination, PDOP is minimized around $45^\circ$--$55^\circ$, indicating
that coverage and PDOP optima may not coincide.

\begin{figure}[ht!]
  \centering
  \begin{minipage}{0.3\textwidth}
    \centering
    \includegraphics[height=50mm,trim={20mm 0mm 20mm 10mm}]
      {images/coverage/orbits_ma9751_inc55.0_walker3x4_hybrid0.pdf}
    \subcaption{S-ELFO Walker $55^{\circ}: 12/3/1$}
  \end{minipage}\hfill
  \begin{minipage}{0.68\textwidth}
    \centering
    \includegraphics[height=50mm]
      {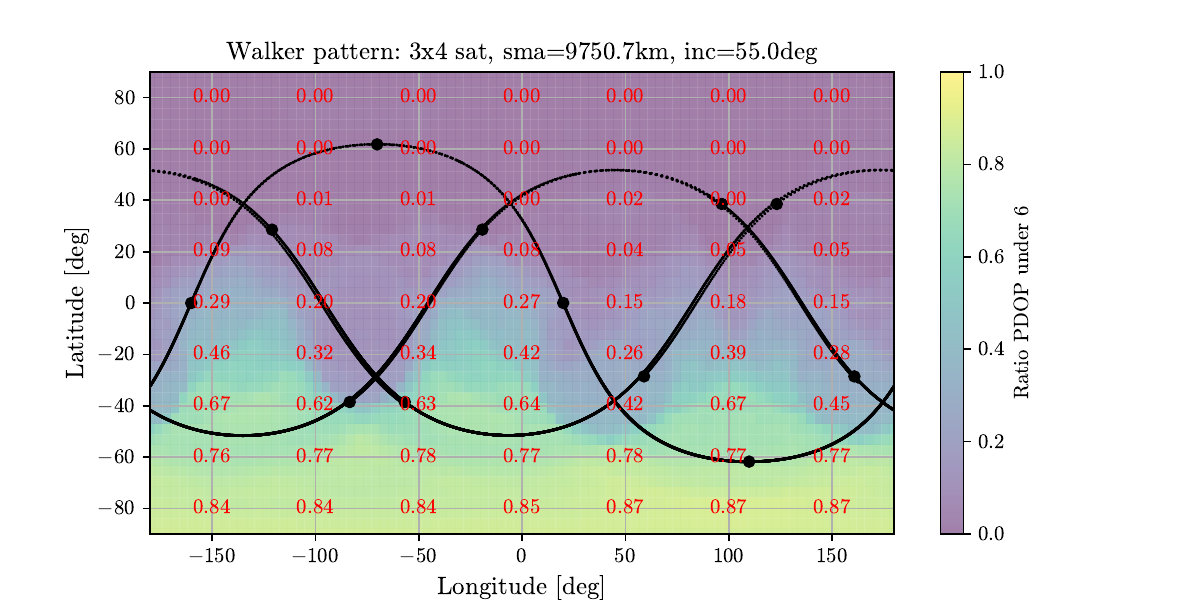}
    \subcaption{Time fraction with PDOP $\leq 6$. Black lines show the ground
    tracks (lunar rotation neglected) over one orbital period.}
  \end{minipage}
  \caption{Orbit and PDOP for S-ELFO Walker $55^{\circ}: 12/4/1$ with
  $a=9751$ km (24-hour orbit).}
  \label{fig:pole_dopmap}
\end{figure}

Figure~\ref{fig:pole_dopmap} maps the fraction of time with PDOP $\leq 6$ for
S-ELFO $55^\circ\!:\, 12/4/1$ at $a=9751$ km. Most of the South Pole region
meets PDOP $\leq 6$, whereas mid- to high-latitude areas in the northern
hemisphere generally do not.

\subsection{North-South ELFO Walker Constellation}
S-ELFO excels at South Pole coverage but performs poorly in the northern
hemisphere. To balance global performance, NS-ELFO splits the argument of periapsis $\omega$ between
$90^\circ$ and $270^\circ$.

\begin{figure}[ht!]
  \centering
  \includegraphics[width=\textwidth]
    {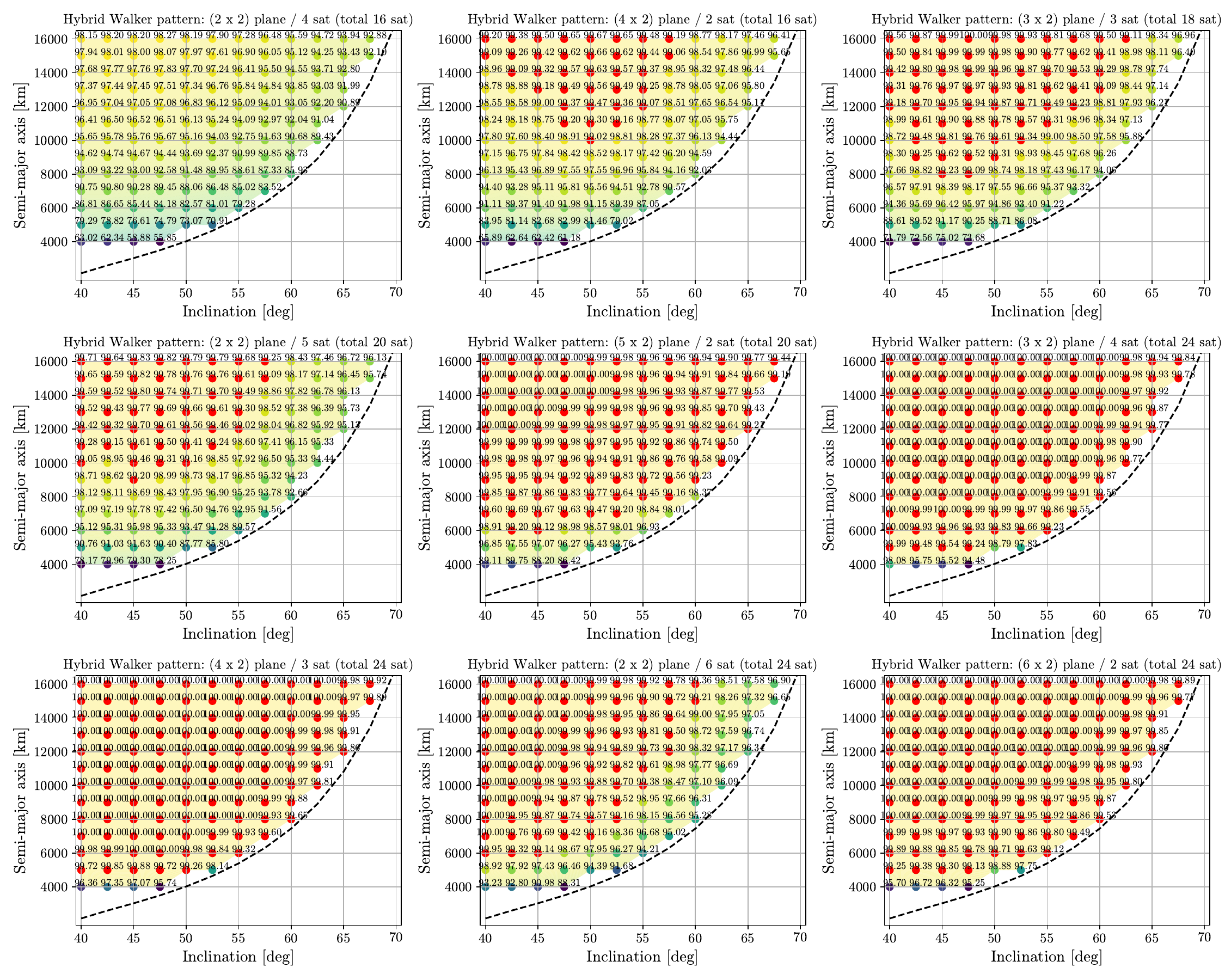}
  \caption{Global coverage for NS-ELFO vs.\ $a$ and $i^{op}$ across
  $(N_p,N_{spp})$. White regions are infeasible.}
  \label{fig:hybrid_coverage}
\end{figure}

Figure~\ref{fig:hybrid_coverage} shows global coverage for NS-ELFO across
$(N_p,N_{spp})$. The minimum to achieve $\geq 99\%$ global 4-fold coverage is
16 satellites. Coverage improves with $a$ for all cases. For $N_s<20$, the
best performance typically occurs at $i^{op}\in[45^\circ,50^\circ]$, enabling
4-fold coverage with smaller $a$.

\begin{figure}[ht!]
  \centering
  \includegraphics[width=\textwidth]
    {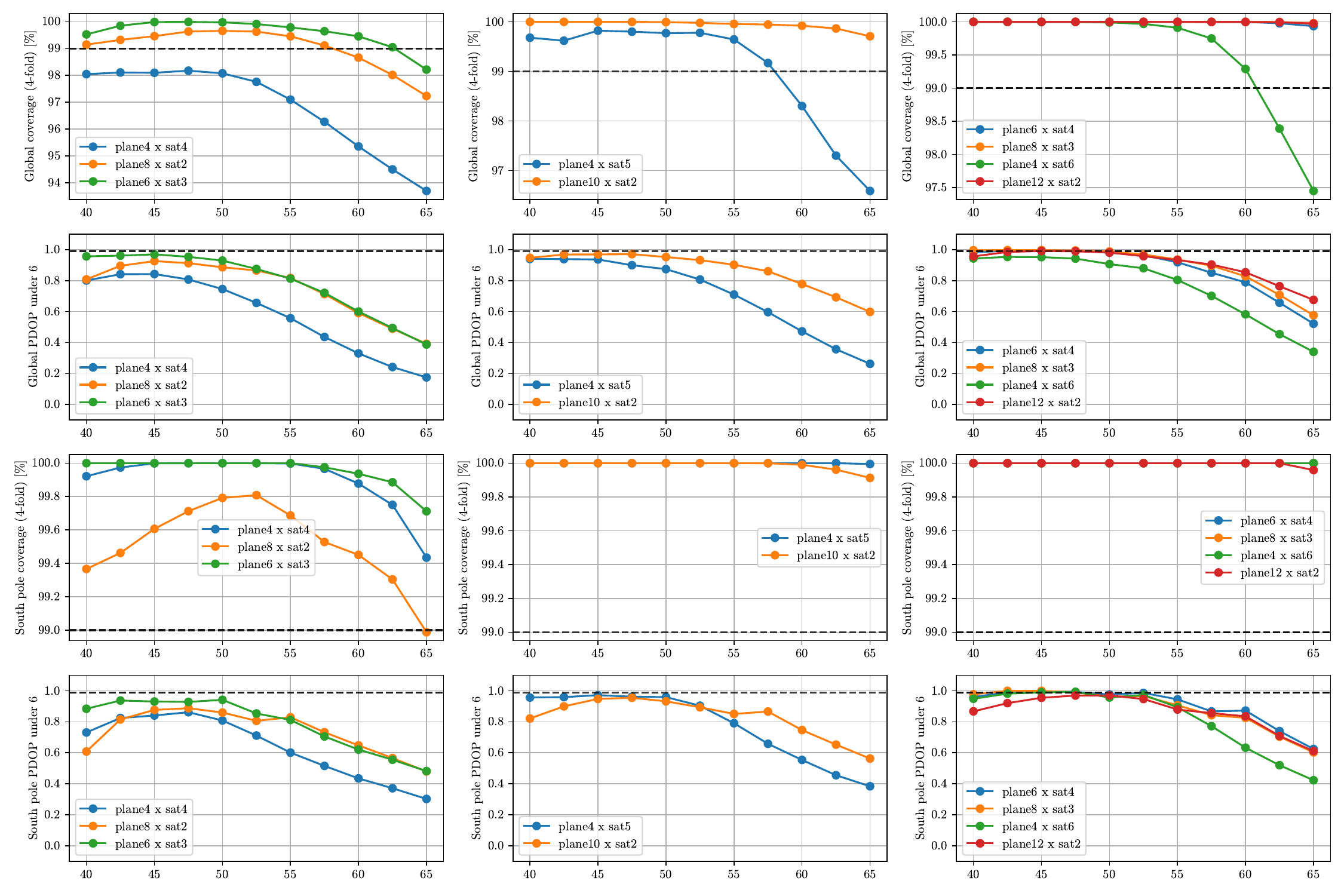}
  \caption{Coverage and PDOP for NS-ELFO at $a=15478$ km. Rows (top to
  bottom): global coverage, global PDOP $\leq 6$ fraction, South Pole
  coverage, South Pole PDOP $\leq 6$ fraction. Columns: $N_s\in \{16,18\}$,
  $N_s=20$, and $N_s=24$.}
  \label{fig:hybrid_pdop}
\end{figure}

Figure~\ref{fig:hybrid_pdop} shows the global and south-pole region coverage for NS-ELFO 
constellation families with semi-major axis fixed to $a=15478$ km. Global 4-fold coverage is
achieved with a minimum of 18 satellites (6 planes or 8 planes), and PDOP $\leq 6$ is satisfied globally with a minimum of 24 satellites.

\begin{figure}[ht!]
  \centering
  \begin{minipage}{0.3\textwidth}
    \centering
    \includegraphics[height=50mm,trim={20mm 0mm 20mm 10mm}]
      {images/coverage/orbits_ma9751_inc50.0_walker3x4_hybrid1.pdf}
    \subcaption{NS-ELFO Walker $50^{\circ}: 24/6/1$}
  \end{minipage}\hfill
  \begin{minipage}{0.68\textwidth}
    \centering
    \includegraphics[height=50mm]
      {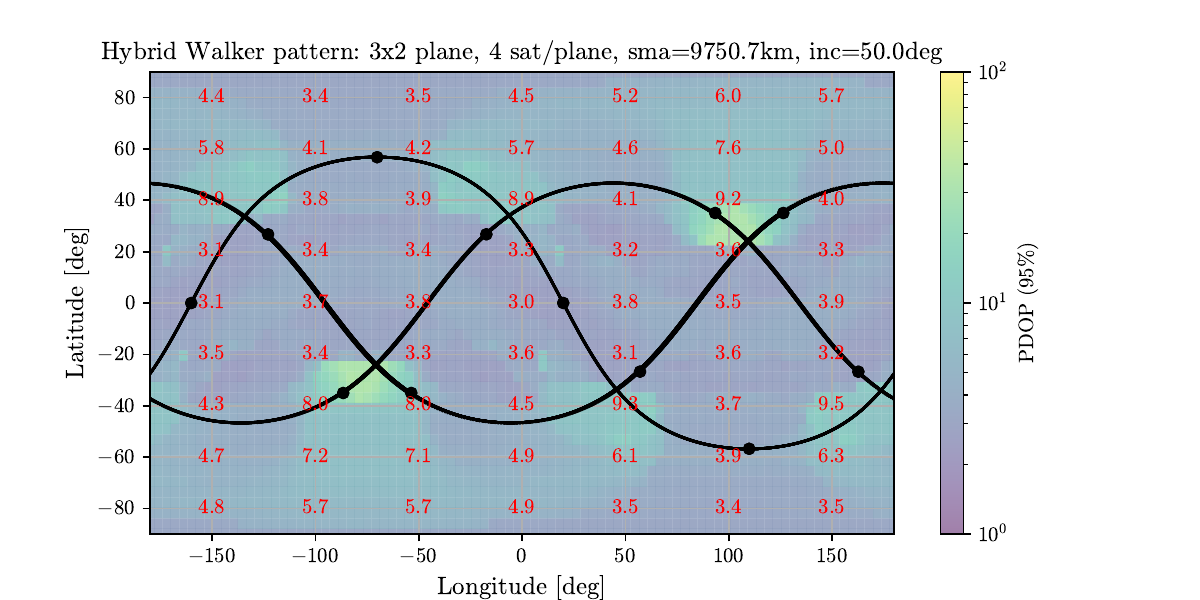}
    \subcaption{Spatial distribution of 95th-percentile PDOP. Black lines show
    ground tracks (no lunar rotation) over one orbital period; red values annotate PDOP.}
  \end{minipage}
  \caption{Orbit and PDOP for NS-ELFO Walker $50^{\circ}: 24/6/1$ with
  $a=9751$ km (24-hour orbit).}
  \label{fig:hybrid_dopmap}
\end{figure}

Figure~\ref{fig:hybrid_dopmap} shows the spatial distribution of the
95th-percentile PDOP for NS-ELFO $50^\circ\!:\, 24/6/1$ at $a=9751$ km. The
constellation provides balanced global PDOP, with elevated values ($> 7$)
at some mid-latitudes.

\subsection{CLFO Walker Constellation}

\begin{figure}[ht!]
  \centering
  \includegraphics[width=\textwidth]
    {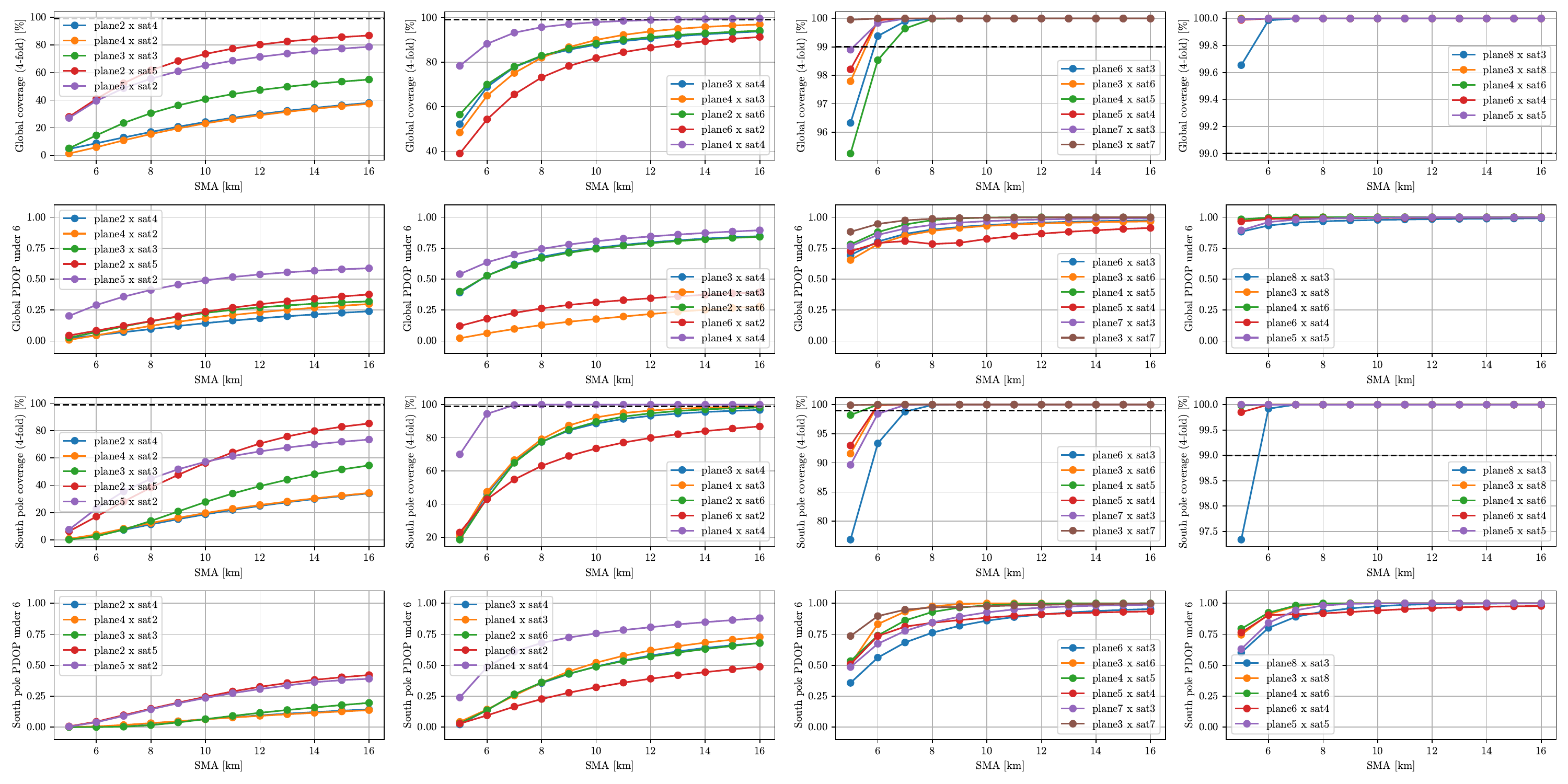}
  \caption{Coverage and PDOP for CLFO with $i^{op}=39.2^\circ$ and
  $a\in[5000,16000]$ km. Rows (top to bottom): global coverage, global PDOP
  $\leq 6$ fraction, South Pole coverage, South Pole PDOP $\leq 6$ fraction.
  Columns (left to right): $N_s\in \{8,9,10\}$, $\{12,16\}$, $\{18,20,21\}$, $\{24,25\}$.}
  \label{fig:circular_pdop}
\end{figure}

Figure~\ref{fig:circular_pdop} summarizes coverage and PDOP for CLFO across
$(N_p,N_{spp})$. As with NS-ELFO, $\geq 99\%$ global 4-fold coverage is
achieved with $N_s=16$ (four planes). With $N_s\geq 20$, both global and
South Pole coverage reaches $\geq 99\%$ with PDOP $\leq 6$.

Because $i^{op}=39.23^\circ$, CLFO underperforms S-ELFO at high
southern latitudes when $N_s$ is small. For example, with $N_s=10$, CLFO
delivers only $\sim 50\%$ South Pole coverage at $a=9{,}000$--$10{,}000$ km,
whereas S-ELFO can exceed $90\%$ (and PDOP $\leq 6$ over $>75\%$ of the time)
with 10 satellites (Figure~\ref{fig:pole_pdop}).

\begin{figure}[ht!]
  \centering
  \begin{minipage}{0.3\textwidth}
    \centering
    \includegraphics[height=50mm,trim={20mm 0mm 20mm 10mm}]
      {images/coverage/orbits_ma9751_inc39.2_walker4x6_hybrid0.pdf}
    \subcaption{CLFO Walker $39.2^{\circ}: 24/4/1$}
  \end{minipage}\hfill
  \begin{minipage}{0.68\textwidth}
    \centering
    \includegraphics[height=50mm]
      {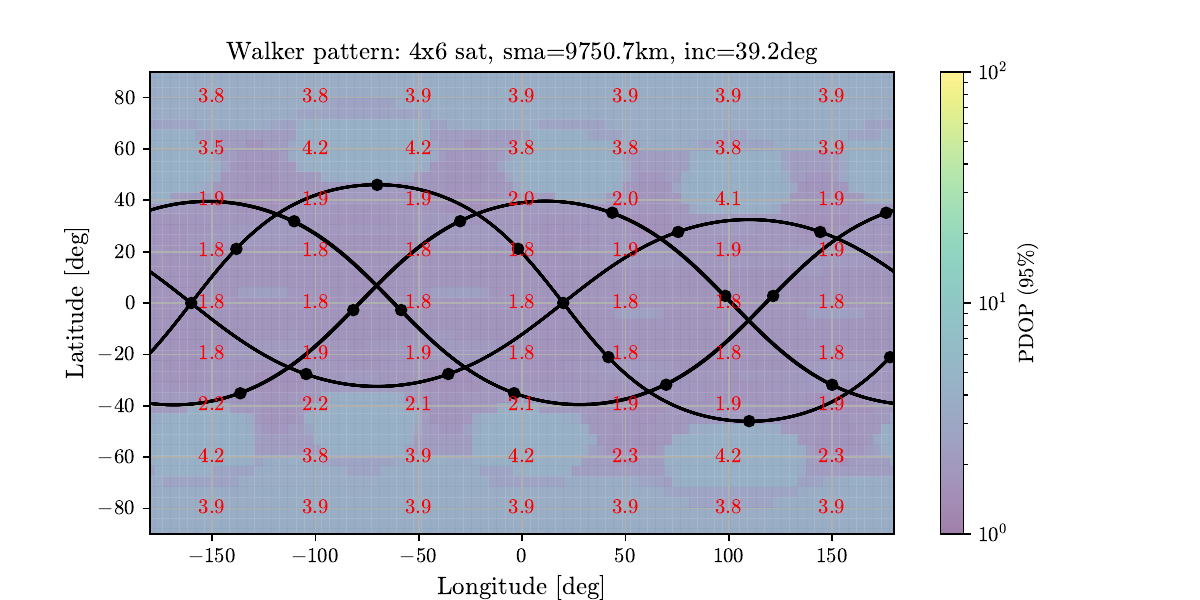}
    \subcaption{Spatial distribution of 95th-percentile PDOP. Black lines show
    ground tracks (no lunar rotation) for one orbital period; red annotations
    indicate PDOP.}
  \end{minipage}
  \caption{Orbit and PDOP for CLFO Walker $39.2^{\circ}: 24/4/1$ with
  $a=9751$ km (24-hour orbit).}
  \label{fig:circular_dopmap}
\end{figure}

Figure~\ref{fig:circular_dopmap} shows the spatial distribution of the
95th-percentile PDOP for CLFO $39.2^\circ\!:\, 24/4/1$ at $a=9751$ km. PDOP
is excellent at low to mid latitudes, with higher values ($> 4$) at
high latitudes.

\subsection{Robustness to Single-Satellite Faults}
\label{sec:fault_robustness}
To assess robustness to single-satellite failures, we repeat the coverage and
PDOP analyses assuming one satellite is non-operational. For each
configuration, we simulate the failure of a satellite and evaluate coverage and
PDOP with the remaining $N_s-1$ satellites. Due to the symmetry of the Walker
constellations, we only need to simulate the failure of one satellite per plane.

The results for S-ELFO, NS-ELFO, and CLFO are summarized in
Tables~\ref{tab:polar_fault}, \ref{tab:hybrid_fault}, and
\ref{tab:clfo_fault}, respectively. 
As expected, the impact of a single-satellite fault is more pronounced when the total number of satellites is small, 
and for metrics that are more difficult to satisfy (e.g., South Pole PDOP $\leq 3$).
For 4-fold coverage and PDOP $\leq 6$ ratio, there is minor or none degradation with $N_s \geq 12$ for S-ELFO (for south-pole coverage) and $N_s \geq 21$ for NS-ELFO and CLFO (for global coverage).

For CLFO, distributing satellites across more planes (fewer per plane) generally improves positioning performance and fault robustness when  $\geq$ 20.
For example, with 24 satellites, the $3$-plane configuration outperforms the $4, 6, 8$-plane configuration in all metrics, and and suffers less degradation under a single-satellite fault.

\begin{table}[htb!]
\caption{The Impact of Single-satellite Fault on South Pole Coverage and PDOP for S-ELFO Constellations}
\label{tab:polar_fault}
\begin{tblr}{
  colspec={X[c] X[c] | X[c]X[c]X[c] |  X[c]X[c]X[c] | X[c]X[c]X[c]},
  width=\textwidth,
  row{even} = {white, font=\small},
  row{odd} = {bg=black!10, font=\small},
  row{1, 2} = {bg=black!20, font=\bfseries\small},
  hline{Z} = {1pt, solid, black!60},
  rowsep=3pt
}
\SetCell[r=2]{c} \textbf{Number of Plane} & 
\SetCell[r=2]{c} \textbf{Number of Sats} &  
\SetCell[c=3]{c} \textbf{South-Pole 4-fold Coverage (\%)} & & &
\SetCell[c=3]{c} \textbf{South-Pole PDOP $\leq 6$ (\%)} & & &
\SetCell[c=3]{c} \textbf{South-Pole PDOP $\leq 3$ (\%)} & & \\
 & & No Fault & Fault & Difference & No Fault & Fault & Difference & No Fault & Fault & Difference \\
\hline
   2 &  6   &    53.43 &  20.13 & 33.30    &     28.37 &  14.67 &  13.69    &      4.90 &   0.33 &   4.58 \\ 
   3 &  6   &    84.36 &  81.95 &  2.41    &     44.69 &  36.92 &   7.77    &     13.56 &   7.22 &   6.34 \\ 
   2 &  8   &    88.60 &  85.37 &  3.23    &     72.77 &  60.25 &  12.52    &     22.32 &  15.12 &   7.20 \\ 
   4 &  8   &    90.28 &  86.59 &  3.69    &     69.35 &  59.94 &   9.40    &     28.74 &  15.41 &  13.33 \\ 
   3 &  9   &    95.01 &  93.91 &  1.11    &     80.07 &  70.66 &   9.41    &     38.80 &  27.08 &  11.72 \\ 
   2 & 10   &    98.51 &  98.51 &  0.00    &     85.72 &  82.21 &   3.52    &     45.26 &  39.36 &   5.89 \\ 
   5 & 10   &    96.25 &  95.70 &  0.55    &     87.38 &  82.72 &   4.67    &     45.75 &  34.63 &  11.12 \\ 
   3 & 12   &   100.00 & 100.00 &  0.00    &     96.67 &  95.95 &   0.72    &     69.65 &  60.63 &   9.02 \\ 
   4 & 12   &   100.00 & 100.00 &  0.00    &     95.20 &  93.84 &   1.36    &     68.62 &  61.24 &   7.38 \\ 
   2 & 12   &   100.00 & 100.00 &  0.00    &     90.20 &  90.10 &   0.10    &     63.71 &  58.68 &   5.03 \\ 
   6 & 12   &    98.50 &  98.49 &  0.01    &     91.47 &  90.09 &   1.38    &     56.47 &  48.95 &   7.53 \\ 
   2 & 14   &   100.00 & 100.00 &  0.00    &     98.24 &  98.17 &   0.07    &     75.47 &  69.91 &   5.56 \\ 
   4 & 16   &   100.00 & 100.00 &  0.00    &     99.49 &  99.37 &   0.11    &     90.20 &  86.61 &   3.59 \\ 
   2 & 16   &   100.00 & 100.00 &  0.00    &     99.92 &  99.91 &   0.01    &     84.86 &  82.65 &   2.21 \\ 
   8 & 16   &   100.00 & 100.00 &  0.00    &     94.10 &  93.99 &   0.11    &     74.02 &  68.76 &   5.26 \\ 
\end{tblr}
\end{table}

\begin{table}[htb!]
\caption{The Impact of Single-satellite Fault on Global Coverage and PDOP for NS-ELFO Constellations}
\label{tab:hybrid_fault}
\begin{tblr}{
  colspec={X[c] X[c] | X[c]X[c]X[c] |  X[c]X[c]X[c] | X[c]X[c]X[c]},
  width=\textwidth,
  row{even} = {white, font=\small},
  row{odd} = {bg=black!10, font=\small},
  row{1, 2} = {bg=black!20, font=\bfseries\small},
  hline{Z} = {1pt, solid, black!60},
  rowsep=3pt
}
\SetCell[r=2]{c} \textbf{Number of Plane} & 
\SetCell[r=2]{c} \textbf{Number of Sats} &  
\SetCell[c=3]{c} \textbf{Global 4-fold Coverage (\%)} & & &
\SetCell[c=3]{c} \textbf{Global PDOP $\leq 6$ (\%)} & & &
\SetCell[c=3]{c} \textbf{Global PDOP $\leq 3$ (\%)} & & \\
 & & No Fault & Fault & Difference & No Fault & Fault & Difference & No Fault & Fault & Difference \\
\hline
   4 & 16   &    98.18 &  96.84 &  1.33    &     80.71 &  77.50 &   3.21    &     42.50 &  35.67 &   6.83 \\ 
   8 & 16   &    99.46 &  98.52 &  0.95    &     92.57 &  88.00 &   4.57    &     47.29 &  37.32 &   9.98 \\ 
   6 & 18   &    99.85 &  99.68 &  0.18    &     96.10 &  92.86 &   3.24    &     64.77 &  51.92 &  12.85 \\ 
   4 & 20   &    99.80 &  99.41 &  0.40    &     89.93 &  89.10 &   0.82    &     66.94 &  60.45 &   6.49 \\ 
  10 & 20   &   100.00 &  99.93 &  0.07    &     96.87 &  95.60 &   1.27    &     67.40 &  61.85 &   5.55 \\ 
   6 & 24   &   100.00 &  99.99 &  0.01    &     99.57 &  99.30 &   0.27    &     86.76 &  79.54 &   7.22 \\ 
   8 & 24   &   100.00 & 100.00 &  0.00    &     99.67 &  99.53 &   0.14    &     85.91 &  82.90 &   3.01 \\ 
   4 & 24   &   100.00 &  99.89 &  0.11    &     95.26 &  95.11 &   0.15    &     83.63 &  80.25 &   3.39 \\ 
  12 & 24   &   100.00 & 100.00 &  0.00    &     99.08 &  98.89 &   0.19    &     76.30 &  71.86 &   4.44 \\ 
\end{tblr}
\end{table}

\begin{table}[htb!]
\caption{The Impact of Single-satellite Fault on Global Coverage and PDOP for CLFO Constellations}
\label{tab:clfo_fault}
\begin{tblr}{
  colspec={X[c] X[c] | X[c]X[c]X[c] |  X[c]X[c]X[c] | X[c]X[c]X[c]},
  width=\textwidth,
  row{even} = {white, font=\small},
  row{odd} = {bg=black!10, font=\small},
  row{1, 2} = {bg=black!20, font=\bfseries\small},
  hline{Z} = {1pt, solid, black!60},
  rowsep=3pt
}
\SetCell[r=2]{c} \textbf{Number of Plane} & 
\SetCell[r=2]{c} \textbf{Number of Sats} &  
\SetCell[c=3]{c} \textbf{Global 4-fold Coverage (\%)} & & &
\SetCell[c=3]{c} \textbf{Global PDOP $\leq 6$ (\%)} & & &
\SetCell[c=3]{c} \textbf{Global PDOP $\leq 3$ (\%)} & & \\
 & & No Fault & Fault & Difference & No Fault & Fault & Difference & No Fault & Fault & Difference \\
\hline
   2 &  8   &    38.05 &  19.01 & 19.04    &     23.98 &  11.99 &  11.99    &      2.12 &   0.87 &   1.25 \\ 
   4 &  8   &    37.41 &  18.65 & 18.77    &     29.84 &  14.87 &  14.97    &      1.94 &   0.94 &   1.00 \\ 
   3 &  9   &    54.89 &  36.29 & 18.60    &     31.96 &  20.14 &  11.82    &      9.59 &   4.97 &   4.63 \\ 
   2 & 10   &    86.71 &  58.92 & 27.79    &     37.56 &  27.06 &  10.50    &     14.72 &   9.79 &   4.93 \\ 
   5 & 10   &    78.58 &  58.88 & 19.70    &     58.74 &  41.57 &  17.17    &      8.23 &   5.04 &   3.19 \\ 
   3 & 12   &    93.78 &  86.03 &  7.74    &     84.65 &  72.26 &  12.39    &     16.58 &   9.76 &   6.82 \\ 
   4 & 12   &    96.91 &  82.54 & 14.37    &     27.85 &  21.31 &   6.54    &     23.30 &  17.75 &   5.55 \\ 
   2 & 12   &    94.09 &  86.46 &  7.64    &     84.17 &  73.43 &  10.74    &     24.83 &  17.15 &   7.67 \\ 
   6 & 12   &    91.31 &  82.69 &  8.62    &     39.39 &  29.43 &   9.96    &     16.08 &  11.07 &   5.02 \\ 
   4 & 16   &    99.58 &  98.07 &  1.51    &     89.44 &  85.63 &   3.81    &     63.52 &  54.06 &   9.47 \\ 
   6 & 18   &   100.00 &  99.99 &  0.01    &     97.44 &  95.63 &   1.81    &     49.48 &  43.82 &   5.67 \\ 
   3 & 18   &   100.00 &  99.92 &  0.08    &     96.59 &  94.85 &   1.74    &     64.95 &  58.60 &   6.35 \\ 
   4 & 20   &   100.00 & 100.00 &  0.00    &     99.91 &  98.97 &   0.94    &     76.05 &  63.73 &  12.33 \\ 
   5 & 20   &   100.00 & 100.00 &  0.00    &     91.32 &  88.37 &   2.95    &     74.39 &  69.47 &   4.92 \\ 
   7 & 21   &   100.00 & 100.00 &  0.00    &     99.28 &  98.94 &   0.34    &     59.37 &  54.07 &   5.30 \\ 
   3 & 21   &   100.00 & 100.00 &  0.00    &     99.98 &  99.94 &   0.04    &     91.23 &  86.13 &   5.10 \\ 
   8 & 24   &   100.00 & 100.00 &  0.00    &     99.07 &  98.77 &   0.30    &     69.28 &  65.06 &   4.22 \\ 
   3 & 24   &   100.00 & 100.00 &  0.00    &    100.00 & 100.00 &   0.00    &     91.73 &  85.97 &   5.76 \\ 
   4 & 24   &   100.00 & 100.00 &  0.00    &    100.00 &  99.87 &   0.13    &     89.13 &  84.84 &   4.30 \\ 
   6 & 24   &   100.00 & 100.00 &  0.00    &     99.78 &  99.66 &   0.12    &     80.57 &  78.69 &   1.89 \\ 
   5 & 25   &   100.00 & 100.00 &  0.00    &     99.85 &  99.62 &   0.23    &     98.21 &  91.79 &   6.42 \\ 
\end{tblr}
\end{table}

\subsection{Summary}
\label{sec:summary_coverage}

Table~\ref{tab:tradeoff_constellation} summarizes the minimum constellation
sizes needed to meet 4-fold coverage and PDOP $\leq 6$ for $\geq 99\%$ of time
and users, both in the South Pole region (latitude $\leq 70^\circ$) and
globally, for S-ELFO, NS-ELFO, and CLFO. The semi-major axis is limited to
$a \leq 15{,}478$ km (48-hour orbit).

S-ELFO provides the strongest South Pole performance, requiring only 12
satellites for $\geq 99\%$ 4-fold coverage and 16 for PDOP $\leq 6$ over
$\geq 99\%$ of the time, but it performs poorly in the northern hemisphere.
The same minima hold under a single-satellite fault for the South Pole region.

NS-ELFO balances coverage globally, but needs at least 18 satellites for
$\geq 99\%$ global 4-fold coverage and 24 for PDOP $\leq 6$ over $\geq 99\%$.
CLFO provides $\geq 99\%$ global 4-fold coverage with 16 satellites and PDOP
$\leq 6$ with 20 satellites.

A practical deployment strategy is therefore to begin with S-ELFO satellites
to satisfy early South Pole requirements and then augment with North-ELFO or
CLFO satellites to extend coverage globally.

\begin{table}[htb!]
\caption{Minimum Required Number of Satellites to Achieve 4-fold Coverage, PDOP $\leq 6$
for $\geq 99\%$, and PDOP $\leq 3$
for $\geq 90\%$, of the Time and Users (Semi-major axis $a \leq 15{,}478$ km,
48-hour orbit). Bold indicates the smallest count per metric.}
\label{tab:tradeoff_constellation}
\begin{tblr}{
  colspec={X[c] | X[1.5cm, l] |  X[3cm, l] | X[c]X[c]X[c]},
  width=\textwidth,
  row{even} = {white, font=\small},
  row{odd} = {bg=black!10, font=\small},
  row{1} = {bg=black!20, font=\bfseries\small},
  hline{Z} = {1pt, solid, black!60},
  rowsep=3pt
}
\textbf{Region} & \textbf{Fault Sat} & \textbf{Metric}  & \textbf{South ELFO} & \textbf{North+South ELFO} & \textbf{CLFO} \\
\SetCell[r=4]{c} \textbf{South Pole (S $\leq 70^{\circ}$)} 
  & \SetCell[r=2]{c} 0 & 99\% 4-fold coverage & \textbf{12} & 16 & \textbf{12} \\
  &  & 99\% PDOP $\leq 6$                     & \textbf{16} & 24 & 18 \\ \hline
  & \SetCell[r=2]{c} 1 & 99\% 4-fold coverage & \textbf{12} & 16 & 16 \\
  &  & 99\% PDOP $\leq 6$                     & \textbf{16} & 24 & 20 \\ \hline
\SetCell[r=6]{c} \textbf{Global} 
 & \SetCell[r=3]{c} 0  & 99\% 4-fold coverage & -- & \textbf{16} & \textbf{16} \\
 & & 99\% PDOP $\leq 6$ & -- & 24 & \textbf{20} \\
 & & 90\% PDOP $\leq 3$ & -- & $>$ 24  &  \textbf{21} \\ \hline
 & \SetCell[r=3]{c} 1 & 99\% 4-fold coverage  & -- & \textbf{18}  & \textbf{18} \\
 & & 99\% PDOP $\leq 6$  & -- &  24 & \textbf{21} \\ 
 & & 90\% PDOP $\leq 3$  & -- &  $>$ 24  & \textbf{25} \\
\end{tblr}
\end{table}

\section{Orbit Determination Error Analysis}
\label{sec:od_analysis}
Orbit determination (OD) error directly impacts the user range error and
is therefore a critical driver of lunar navigation constellation design. 
This section evaluates how orbital elements influence OD performance using covariance analysis.

Several OD approaches have been proposed for lunar navigation satellites,
including Earth-based tracking~\citep{Sestanavi2025}, reception of GNSS sidelobe
signals~\citep{Delepaut2020, Iiyama_navigation_2024}, and inter-satellite links
(ISL)~\citep{cheetham2022capstone}. 
Here we focus on Earth-based tracking, the most established method for deep-space missions.

Time synchronization is not considered in this section, since by using two-way time transfer, the time offset between the satellite and ground station clocks can be estimated in a decoupled manner from the orbit estimation~\citep{Sestanavi2025} at
a very high accuracy (sub nano-second level), and does not depend on the orbit elements.

\subsection{Measurement Modeling}
We assume each satellite is tracked by three Lunar Exploration Ground Station
(LEGS) sites located at White Sands (USA), Matjiesfontein (South Africa), and
an Australian site~\citep{NASA2023_LEGSBrochure}. Because an official Australian LEGS site has not been publicly specified, we model Canberra (DSN) as a proxy.
Each station provides two-way X-band range and range-rate. 
Site locations and measurement noise assumptions are summarized in Table~\ref{tab:legs}.

\begin{table}[htb]
\caption{LEGS Site Locations and Assumed Measurement Noise
(1$\sigma$)~\citep{NASA2023_LEGSBrochure}.}
\label{tab:legs}
\begin{tblr}{
  colspec={X[l]X[c]X[c]X[l]},
  width=\textwidth,
  row{even} = {white, font=\small},
  row{odd} = {bg=black!10, font=\small},
  row{1} = {bg=black!20, font=\bfseries\small},
  hline{Z} = {1pt, solid, black!60},
  rowsep=3pt
}
\textbf{Site} & \textbf{Latitude [deg]} & \textbf{Longitude [deg]} & \textbf{Noise (Range / Range-rate, 1$\sigma$)} \\
White Sands (USA) & 32.54 N & 106.61 W & 1 m / 0.1 mm\,s$^{-1}$ \\
Matjiesfontein (South Africa) & 32.23 S & 20.58 E & 1 m / 0.1 mm\,s$^{-1}$ \\
Canberra (Australia) & 35.40 S & 148.98 E & 1 m / 0.1 mm\,s$^{-1}$ \\
\end{tblr}
\end{table}

Range and range-rate are sampled every 1 minute at each site whenever the
elevation angle of the site to satellite direction exceeds $10^\circ$. Measurement noise is zero-mean white Gaussian.
States are estimated in the Moon-Centered Inertial (MCI) frame (origin at the
Moon’s center; axes aligned with J2000).

Let $\mathbf{X}_s = \begin{bmatrix}\mathbf{r}_s & \mathbf{v}_s\end{bmatrix}^\top$
denote the satellite state, and
$\mathbf{X}_{g,i} = \begin{bmatrix}\mathbf{r}_{g,i} & \mathbf{v}_{g,i}\end{bmatrix}^\top$
the state of ground station $i$ (ECEF). The time-varying transform from MCI to
ECEF is
\begin{equation}
  \mathbf{X}^{\mathrm{ecef}}(t)
  = M_{mci}^{ecef}(t)\,\mathbf{X}^{\mathrm{mci}}(t)
  + \mathbf{b}_{mci}^{ecef}(t),
\end{equation}
with $M_{mci}^{ecef}\in\mathbb{R}^{6\times 6}$ and
$\mathbf{b}_{mci}^{ecef}\in\mathbb{R}^{6\times 1}$.

The range and range-rate from station $i$ to the satellite are
\begin{align}
  \rho_i &= \left\|\mathbf{r}_s^{ecef} - \mathbf{r}_{g,i}\right\| + \epsilon_{\rho}, \quad \epsilon_{\rho} \sim \mathcal{N}(0, \sigma_{\rho}) \\
  \dot{\rho}_i &=
  \frac{\big(\mathbf{r}_s^{ecef} - \mathbf{r}_{g,i}\big) \cdot
        \big(\mathbf{v}_s^{ecef} - \mathbf{v}_{g,i}\big)}
       {\left\|\mathbf{r}_s^{ecef} - \mathbf{r}_{g,i}\right\|}
  + \epsilon_{\dot{\rho}}, \quad \epsilon_{\dot{\rho}} \sim \mathcal{N}(0, \sigma_{\dot{\rho}})
\end{align}
where $\sigma_{\rho} = 1 \ \text{m}, \sigma_{\dot{\rho}} = 0.1 \ \text{mm/s}$ are the range and range-rate measurement noises, respectively, and 
\begin{equation}
\mathbf{X}_s^{ecef}
= \begin{bmatrix}\mathbf{r}_s^{ecef} & \mathbf{v}_s^{ecef}\end{bmatrix}^\top
= M_{mci}^{ecef}(t)\,\mathbf{X}_s + \mathbf{b}_{mci}^{ecef}(t)
\end{equation}
The measurement Jacobian at time $t_k$ is given by
\begin{equation}
\small
\begin{aligned}
  H_{i,s,k}
  &= \begin{bmatrix}
       \dfrac{\partial \rho_i}{\partial \mathbf{X}_s} \\
       \dfrac{\partial \dot{\rho}_i}{\partial \mathbf{X}_s}
     \end{bmatrix}
   = M_{mci}^{ecef}(t_k)^\top
     \begin{bmatrix}
       \dfrac{\partial \rho_{i,k}}{\partial \mathbf{X}_s^{ecef}} \\
       \dfrac{\partial \dot{\rho}_{i,k}}{\partial \mathbf{X}_s^{ecef}}
     \end{bmatrix}
  = M_{mci}^{ecef}(t_k)^\top
     \begin{bmatrix}
       \mathbf{u}_{i,s,k}^\top & \mathbf{0}_{1\times 3} \\
       \dfrac{\mathbf{v}_{s,k}^{ecef\,\top}\!\left(I_3 -
       \mathbf{u}_{i,s,k}\mathbf{u}_{i,s,k}^\top\right)}{\rho_{i,k}}
       & \mathbf{u}_{i,s,k}^\top
     \end{bmatrix},
\end{aligned}
\end{equation}
where
$\mathbf{u}_{i,s,k}
= \dfrac{\mathbf{r}_{s,k}^{ecef} - \mathbf{r}_{g,i}}
        {\left\|\mathbf{r}_{s,k}^{ecef} - \mathbf{r}_{g,i}\right\|}$
is the line-of-sight unit vector.

\subsection{Dynamics Model}
The force model includes lunar gravity (spherical harmonics to degree/order 20)
and third-body accelerations from Earth and Sun:
\begin{equation}
  \dot{\mathbf{X}}_s
  = \begin{bmatrix}
      \mathbf{v}_s \\
      -\dfrac{\mu_M}{\|\mathbf{r}_s\|^3}\,\mathbf{r}_s
      + \mathbf{a}_{\mathrm{sph}} + \mathbf{a}_{3b}
    \end{bmatrix},
\end{equation}
where $\mu_M$ is the lunar gravity constant ($\mu_M = 4902 \ \text{km}^3 \text{s}^{-2}$), $\mathbf{a}_{\mathrm{sph}}$ is the lunar spherical-harmonic acceleration, and $\mathbf{a}_{3b}$ is the third-body
acceleration. The state transition matrix (STM) $\Phi(t_{k+1},t_k)$ is obtained
by integrating the following equation from $t = t_k$ to $t = t_{k+1}$.
\begin{align}
  \dot{\Phi}(t,t_k) &= A(t)\,\Phi(t,t_k), \quad t>t_k, \\
  A(t) &= \left.\dfrac{\partial \dot{\mathbf{X}}_s}{\partial \mathbf{X}_s}\right|_{\mathbf{X}_s(t)},
  \qquad \Phi(t_k,t_k) = I_6.
\end{align}

\subsection{Covariance Analysis Methodology}
The covariance analysis is performed using an Extended Kalman Filter (EKF) framework. The state estimate and its covariance matrix are propagated in time using the dynamics model, and updated whenever a new measurement is available using the measurement model described above. The initial state covariance is assumed to be a diagonal matrix with $100^2$ m$^2$ for position and $1^2$ (m/s)$^2$ for velocity. The initial epoch of the simulation is set to January 1, 2030, 12:00:00 UTC.

The covariance update equations are given by~\citep{Landis2009}
\begin{align}
    P_{k|k-1} &= \Phi(t_k, t_{k-1}) P_{k-1|k-1} \Phi(t_k, t_{k-1})^{\top} + Q_k \\
    K_k &= P_{k|k-1} H_k^{\top} (H_k P_{k|k-1} H_k^{\top} + R_k)^{-1} \\
    P_{k|k} &= (I - K_k H_k) P_{k|k-1}
\end{align}
where $K_k$ is the Kalman gain, $P_{k|k-1}$ is the predicted covariance, $P_{k|k}$ is the updated covariance, $H_k$ is the measurement Jacobian matrix, and $R_k$ is the measurement noise covariance matrix computed from $\sigma_{\rho}, \sigma_{\dot{\rho}}$, and $Q_k$ is the process noise covariance matrix.

The process noise covariance $Q_k$ is modeled using a 1st-order Taylor series Truncation as~\citep{CarpenterDSouza2025}
\begin{equation}
    Q_k = \begin{bmatrix}
        \frac{\Delta t^3}{3} I_3 & \frac{\Delta t^2}{2} I_3 \\
        \frac{\Delta t^2}{2} I_3 & \Delta t I_3
    \end{bmatrix} \sigma_a^2
\end{equation}
where $\Delta t = 60 s$ is the time step, and $\sigma_a$ is the standard deviation of the unmodeled acceleration, which is set to $1 \times 10^{-7} \ \text{km} \cdot\text{s}^{-3/2}$ in this analysis.

When creating the Ephemeris to be broadcast to the users (which will be used by users to compute satellite positions), the predicted covariance is used. Therefore, we evaluate the (worst) predicted position error covariance at the time of ephemeris generation, which we assume occurs every 2 hours. The predicted state error covariance is obtained by propagating the updated covariance at each timestep for 2 hours using the state transition matrix.
\begin{eqnarray}
    P_{pred|k} = \Phi(t_{pred}, t_k) P_{k|k} \Phi(t_{pred}, t_k)^{\top}
\end{eqnarray}
where $t_{pred} = t_k + 2$ hours. We do not model the ephemeris generation errors in this analysis for simplicity.

In reality, the OD accuracy is affected by additional perturbations such as solar radiation pressure and spacecraft maneuvers, and measurement biases, such as tropospheric and ionospheric delays, antenna phase center variations, and ground station location errors. These effects are not considered in this analysis for simplicity, but the results presented here provide a baseline for comparing the OD performance of different orbit configurations.

\subsection{Covariance Analysis Results}
We sweep semi-major axis $a\in[4000,17000]$ km (1000 km steps), inclination
$i\in[30^\circ,90^\circ]$ ($10^\circ$ steps), and RAAN
$\Omega\in\{0^\circ,\dots,330^\circ\}$ ($30^\circ$ steps). Eccentricity is set
by \eqref{eq:frozen_condition} and we fix $\omega$ to $90^\circ$. The initial mean anomaly is $0^\circ$. Each case runs for four orbital periods, and we record the 95th
percentile of the predicted position error at the final period.

\begin{figure}
    \centering
        \includegraphics[width=0.6\columnwidth]{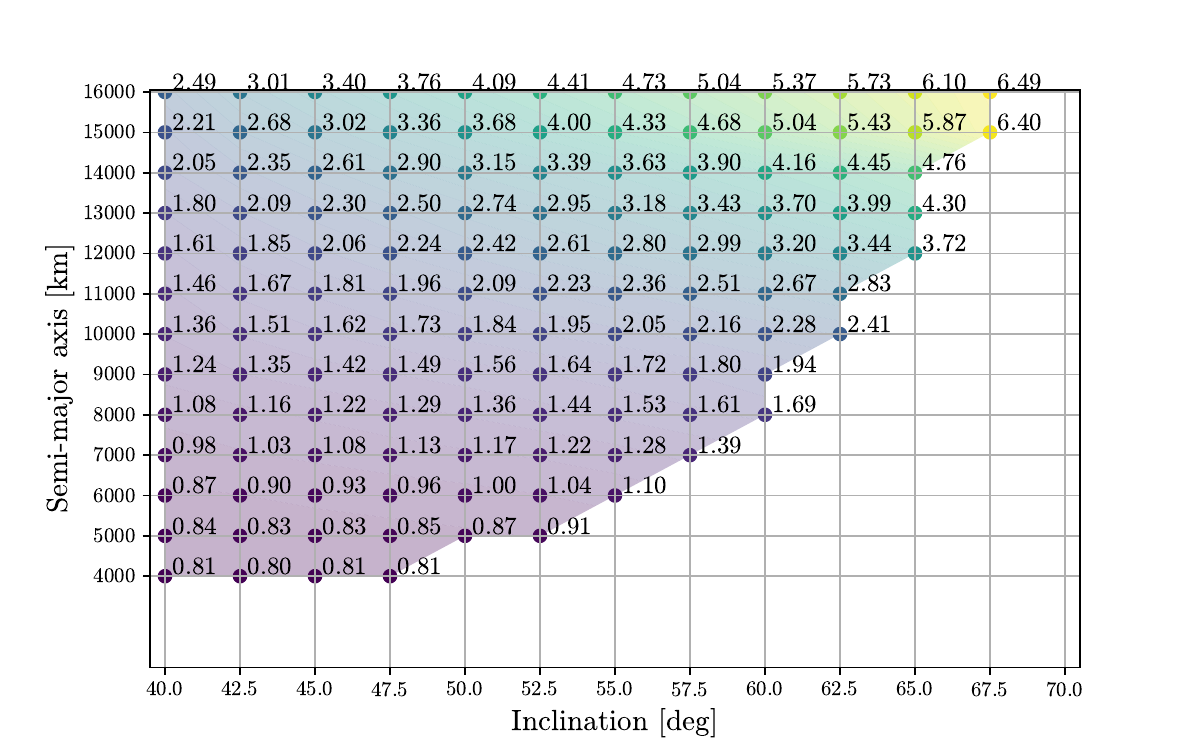}
        \caption{95 percentile of the predicted position error at the time of Ephemeris generation (every 2 hours) for different combinations of semi-major axis $a$ and inclination $i$. The eccentricity is fixed to the frozen orbit condition, and the argument of periapsis is fixed to $\omega = 90^{\circ}$. The RAAN is fixed to $\Omega_0 = 75^{\circ}$.}
        \label{fig:od_sma_inc}
\end{figure}

Figure~\ref{fig:od_sma_inc} shows the 95th-percentile predicted position error
at ephemeris times (every 2 h) vs.\ $a$ and $i$ for $\Omega=75^\circ$. In
general, OD accuracy degrades as $a$ increases, because slower orbital motion
reduces range-rate observability. Lower inclinations tend to perform better, largely because they imply smaller eccentricity and a closer apolune.

\begin{figure}[ht!]
    \centering
    \begin{minipage}
        [b]{0.54\columnwidth}
        \centering
        \includegraphics[height=60mm]{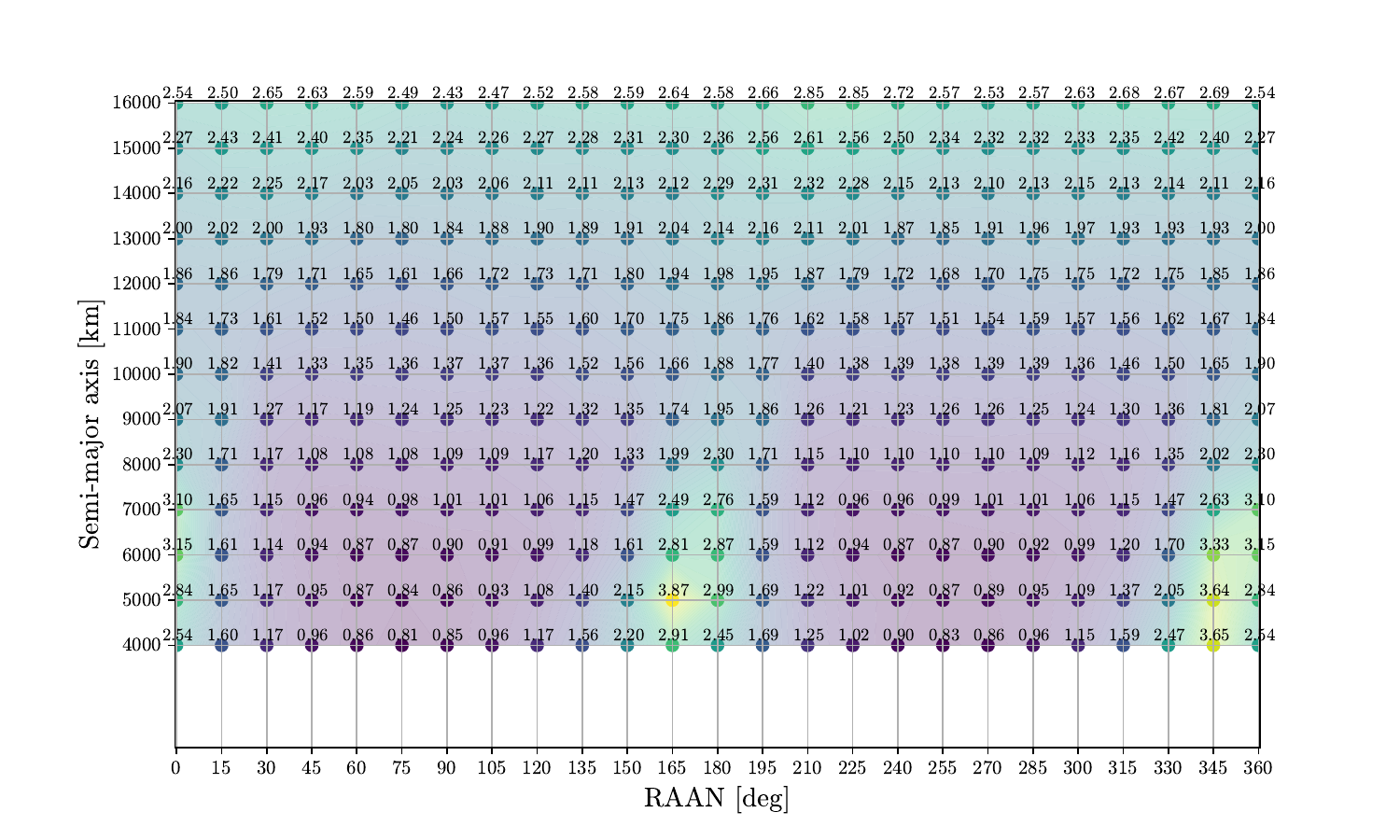}
        \caption{95th percentile of the predicted position error at the time of ephemeris generation (every 2 hours) for different combinations of semi-major axis $a$ and RAAN $\Omega$, with a fixed inclination of $i^{op} = 40$. The eccentricity is fixed to the frozen orbit condition, and the argument of periapsis is fixed to $\omega = 90^{\circ}$.}
        \label{fig:od_sma_omega}
    \end{minipage}
    \hfill
    \begin{minipage}
        [b]{0.44\columnwidth}
        \centering
        \includegraphics[height=60mm, clip={20mm, 30mm, 20mm, 30mm}]{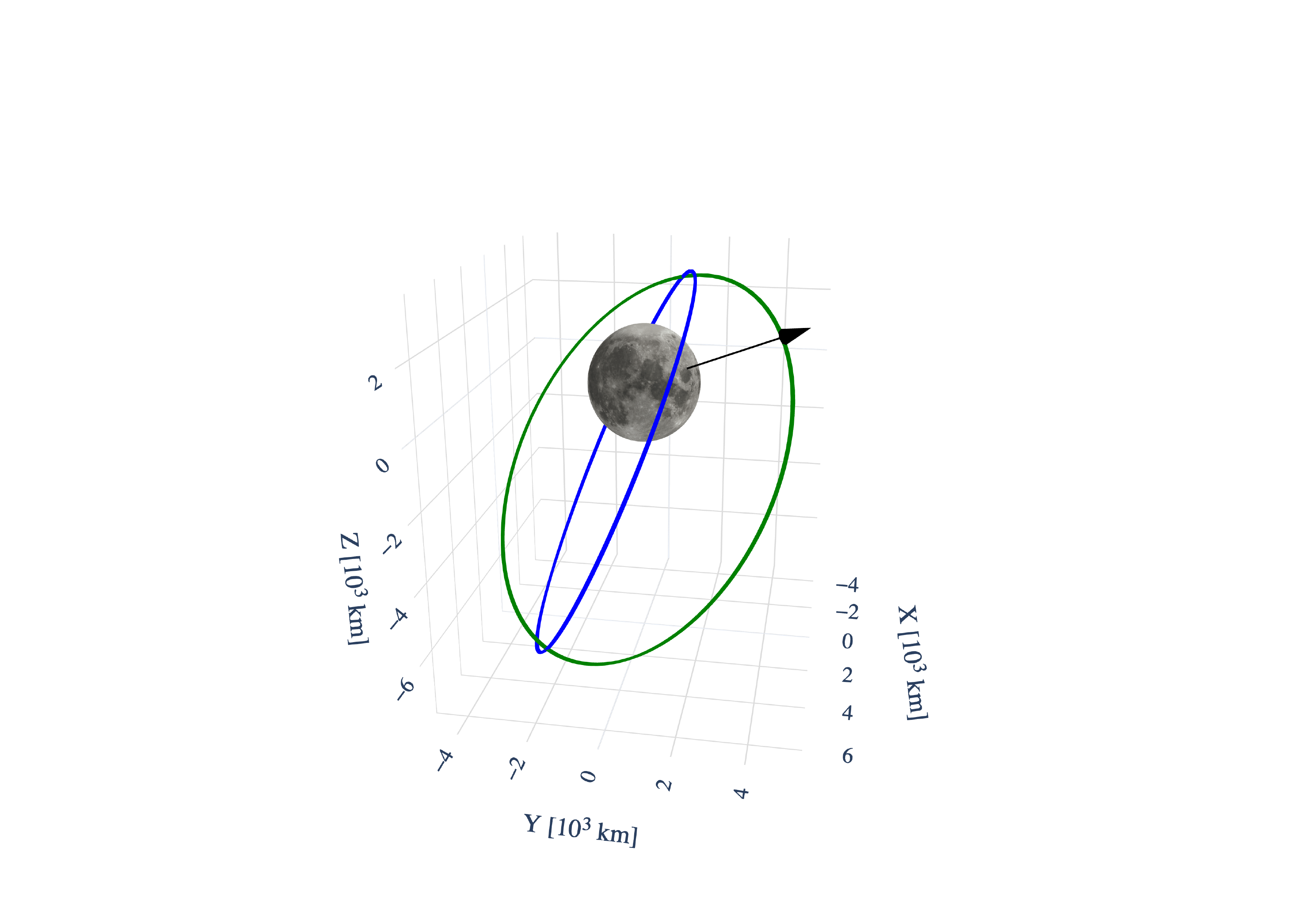}
        \caption{Geometry between the ground stations and the satellite orbit for $a=6000$, $i=45.0$, $e=0.40$, and $\Omega = 70^{\circ}$ (blue) and $\Omega = 165^{\circ}$ (green). When $\Omega = 165^{\circ}$ (green), the satellite orbit is aligned to the line connecting the Earth and Moon (shown in black arrow), which leads to poor observability and a longer occultation period.}
        \label{fig:od_geometry_omega}
    \end{minipage}
\end{figure}

\begin{figure}[ht!]
    \begin{minipage}
        [b]{0.98\columnwidth}
        \centering
        \includegraphics[width=0.75\columnwidth]{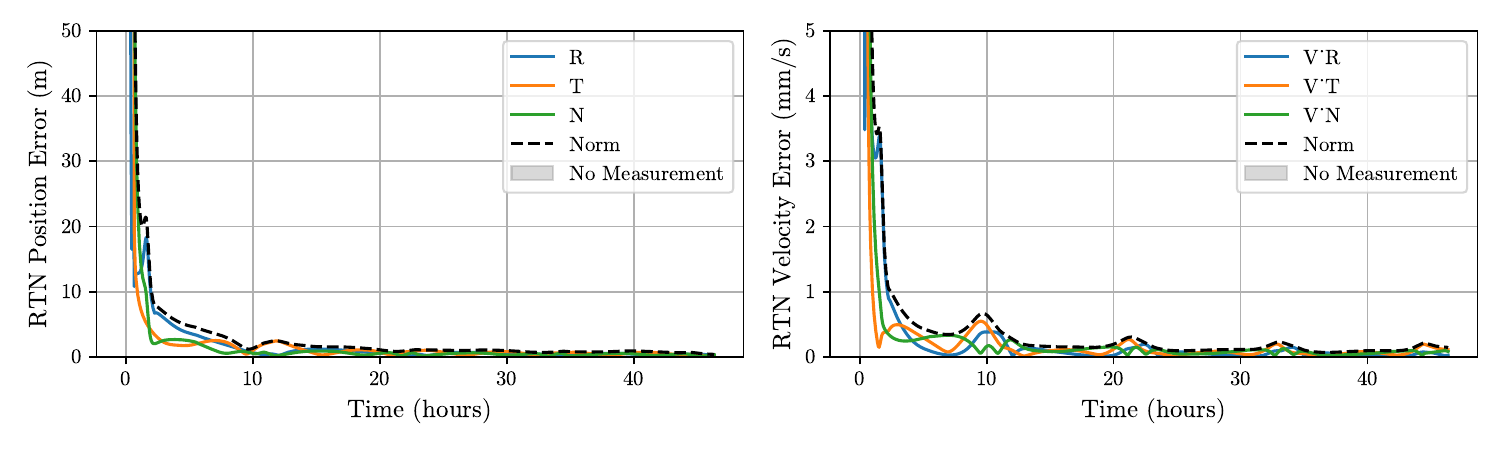}
        \subcaption{OD result for $a=6000$ km, $i=45^{\circ}$, $\Omega = 75^{\circ}$.}
        \label{fig:odsim_omega75}
    \end{minipage}
    \begin{minipage}
        [b]{0.98\columnwidth}
        \centering
        \includegraphics[width=0.75\columnwidth]{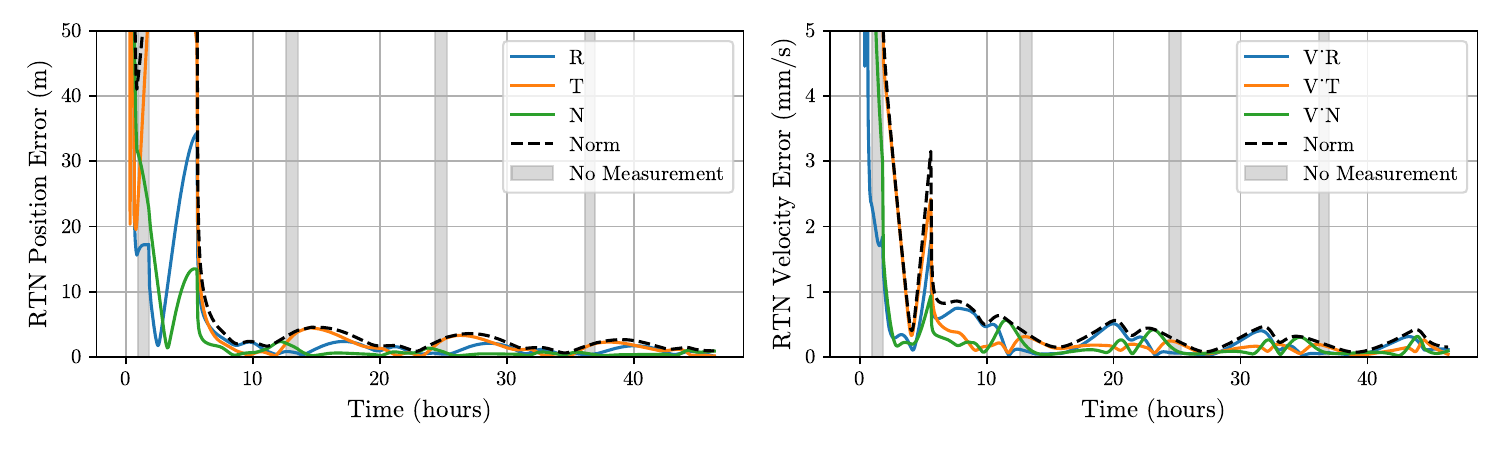}
        \subcaption{OD result for $a=6000$ km, $i=45^{\circ}$, $\Omega = 165^{\circ}$. The gray area shows the area when the satellite is not visible from any ground station.}
        \label{fig:odsim_omega165}
    \end{minipage}
    \caption{Covariance analysis results for two different RAAN values. The top figures show the ground
    tracks of the satellite and the ground stations, while the bottom figures show the position error (solid lines) and the 3-sigma predicted position error (dashed lines). The OD accuracy is significantly worse when $\Omega = 120^{\circ}$ (top) compared to $\Omega_0 = 30^{\circ}$ (bottom) due to poor observability.}
    \label{fig:od_results}
\end{figure}

Figure~\ref{fig:od_sma_omega} shows sensitivity to RAAN for fixed
$i^{op}=40^\circ$. RAAN significantly affects OD accuracy, especially at smaller
$a$, through geometry with respect to the ground network. When
$\Omega=165^\circ$, the orbit is nearly aligned with the Earth–Moon line
(Figure~\ref{fig:od_geometry_omega}), reducing observability and increasing
occultation time. Figure~\ref{fig:od_results} compares full covariance traces
for $\Omega=75^\circ$ and $165^\circ$ at $a=6000$ km, $i=45^\circ$.

\section{Receiver Noise Analysis}
\label{sec:receiver_noise}
The user range error (URE) due to receiver noise is affected not only by the satellites’ orbit determination (OD) accuracy but also by thermal noise in the receiver’s delay lock loop (DLL).
The DLL thermal noise depends on received signal power as follows~\citep{kaplan2005understanding}:
\begin{equation}
\left(\sigma_{\mathrm{DLL} t}^2\right)^{(k)}=
\begin{cases}
\displaystyle \frac{B_{\mathrm{DLL}}}{2\left(C / N_0\right)_t^{(k)}}\, d \left(1+\frac{2}{T\left(C / N_0\right)_t^{(k)}(2-d)}\right), & d \geq \dfrac{\pi}{T_{\mathrm{c}} B_{\mathrm{fe}}} \\\\
\displaystyle \frac{B_{\mathrm{DLL}}}{2\left(C / N_0\right)_t^{(k)}} \left(\frac{1}{T_{\mathrm{c}} B_{\mathrm{fe}}}\right) \left(1+\frac{1}{T\left(C / N_0\right)_t^{(k)}}\right), & d \leq \dfrac{\pi}{T_{\mathrm{c}} B_{\mathrm{fe}}} \\\\
\displaystyle \frac{B_{\mathrm{DLL}}}{2\left(C / N_0\right)_t^{(k)}} \left(\frac{1}{T_{\mathrm{c}} B_{\mathrm{fe}}}+\frac{B_{\mathrm{fe}} T_{\mathrm{c}}}{\pi-1}\left(d-\frac{1}{T_{\mathrm{c}} B_{\mathrm{fe}}}\right)^2\right) \left(1+\frac{2}{T\left(C / N_0\right)_t^{(k)}(2-d)}\right), & \dfrac{1}{T_{\mathrm{c}} B_{\mathrm{fe}}}<d<\dfrac{\pi}{T_{\mathrm{c}} B_{\mathrm{fe}}}.
\end{cases}
\end{equation}
The parameters and their values are given in Table~\ref{tab:dll_params}. Note that the carrier-to-noise ratio $\left(C / N_0\right)_t^{(k)}$ term in the above equation is in linear scale (not dB-Hz).

The $\left(C / N_0\right)_t^{(k)}$ [dB-Hz] for satellite $k$ at time $t$ is computed from the following link-budget equation:
\begin{equation}
\left(C / N_0\right)_t^{(k)} = P_{\mathrm{tx}} + G_{\mathrm{tx}} + G_{\mathrm{rx}} - L_t^{(k)} - L_{fs}^{(k)} - 10 \log_{10}(k_b T_{\mathrm{sys}}).
\end{equation}
Here $k_b = 1.38 \times 10^{-23}$ J/K is Boltzmann’s constant, and the free-space path loss is
\begin{equation}
L_{fs}^{(k)} = 20 \log_{10}\!\left(\frac{4 \pi d^{(k)} f}{c}\right),
\end{equation}
where $d^{(k)}$ is the user–satellite distance, $f$ is the carrier frequency, and $c$ is the speed of light.
Other parameters and values are listed in Table~\ref{tab:link_budget_params}.

\begin{table}[htb]
\centering
\begin{minipage}[b]{0.49\columnwidth}
\centering
\caption{Parameters for DLL Thermal Noise Calculation}
\label{tab:dll_params}
\begin{tblr}{
  colspec={X[l]X[c]X[c]},
  width=\textwidth,
  row{even} = {white, font=\small},
  row{odd} = {bg=black!10, font=\small},
  row{1} = {bg=black!20, font=\bfseries\small},
  hline{Z} = {1pt, solid, black!60},
  rowsep=3pt
}
\textbf{Parameter} & \textbf{Symbol} & \textbf{Value} \\
Code tracking loop bandwidth & $B_{\mathrm{DLL}}$ & 1.0 Hz \\
Early–late correlator spacing & $d$ & 1.0 chips \\
Coherent integration time & $T$ & 20 ms \\
Chipping period & $T_{\mathrm{c}}$ & 0.196 $\mu$s \\
Double-sided front-end bandwidth & $B_{\mathrm{fe}}$ & 2 MHz \\
\end{tblr}
\end{minipage}
\hfill
\begin{minipage}[b]{0.49\columnwidth}
\centering
\caption{Parameters for Link Budget (Tx: transmitter, Rx: receiver)}
\label{tab:link_budget_params}
\begin{tblr}{
  colspec={X[l]X[c]X[c]},
  width=\textwidth,
  row{even} = {white, font=\small},
  row{odd} = {bg=black!10, font=\small},
  row{1} = {bg=black!20, font=\bfseries\small},
  hline{Z} = {1pt, solid, black!60},
  rowsep=3pt
}
\textbf{Parameter} & \textbf{Symbol} & \textbf{Value} \\
Frequency & $f$ & 2491.005 MHz \\
Tx power & $P_{\mathrm{tx}}$ & 15.0 dBW or 8.0 dBW \\
Tx antenna gain & $G_{\mathrm{tx}}$ & computed \\
Rx antenna gain & $G_{\mathrm{rx}}$ & computed \\
Tx cable loss & $L_t^{(k)}$ & 1 dB \\
Free-space path loss & $L_{fs}^{(k)}$ & computed \\
Rx system temperature & $T_{\mathrm{sys}}$ & 432 K \\
\end{tblr}
\end{minipage}
\end{table}

The satellite transmit antenna pattern follows \citet{Melman2022LCNS} and is scaled by the apparent lunar-horizon angle at apolune:
\begin{equation}
G_{\mathrm{rx}}(\theta) = G_{\mathrm{ref}}(\theta/\alpha), \qquad \alpha = \frac{\tan^{-1}\!\big(\frac{r_{\mathrm{moon}}}{a(1+e)}\big)}{\tan^{-1}\!\big(\frac{r_{\mathrm{moon}}}{a_{\mathrm{ref}}(1+e_{\mathrm{ref}})}\big)}.
\end{equation}
Here $G_{\mathrm{ref}}(\theta)$ is the reference pattern from \citet{Melman2022LCNS}, $r_{\mathrm{moon}}=1737.4$ km is the lunar radius, $a_{\mathrm{ref}}=9750$ km and $e_{\mathrm{ref}}=0.6383$ are reference orbit parameters from \citet{Melman2022LCNS}, and $\theta$ is the off-boresight angle.
Consistent with the OD error analysis, we compute receiver noise URE at each lunar-surface user point every 5 minutes over one orbital period and take the 95th percentile over time as the performance metric.
We also use the receiver antenna pattern $G_{\mathrm{rx}}$ from \citet{Melman2022LCNS}, which has a gain of 6 dBi at zenith.

\begin{figure}[ht!]
\begin{minipage}
[b]{0.49\columnwidth}
    \centering
    \includegraphics[width=\textwidth]{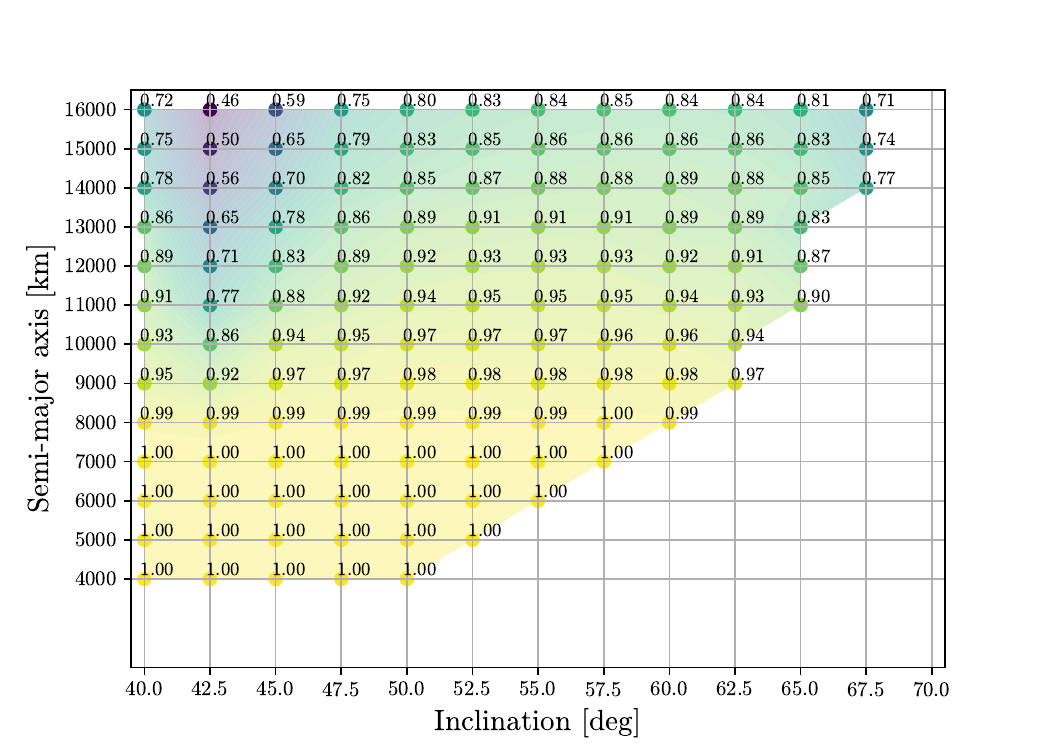}
    \subcaption{Ratio where $C/N_0 \geq $ 30 dB-Hz ($P_{tx}$  = 8 dBW)}
    \label{fig:cn0_Ptx10}
\end{minipage}
\hfill
\begin{minipage}
[b]{0.49\columnwidth}
    \centering
    \includegraphics[width=\textwidth]{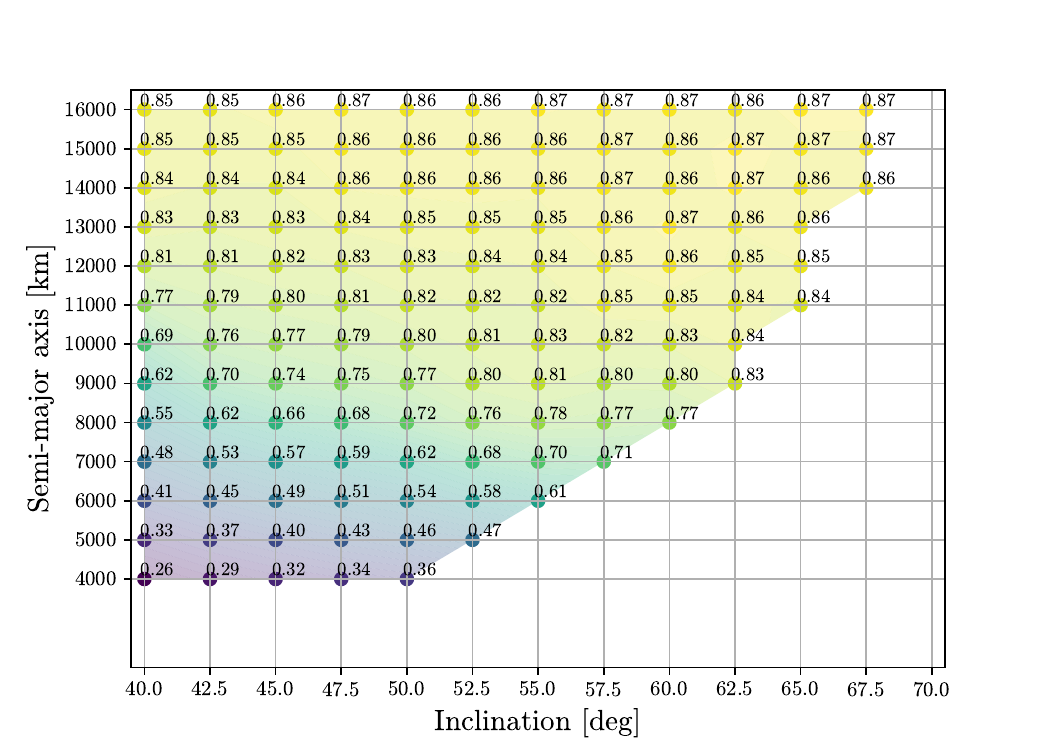}
    \subcaption{95th-percentile URE [m] ($P_{tx}$  = 8 dBW)}
    \label{fig:ure_Ptx10}
\end{minipage}

\begin{minipage}
[b]{0.49\columnwidth}
    \centering
    \includegraphics[width=\textwidth]{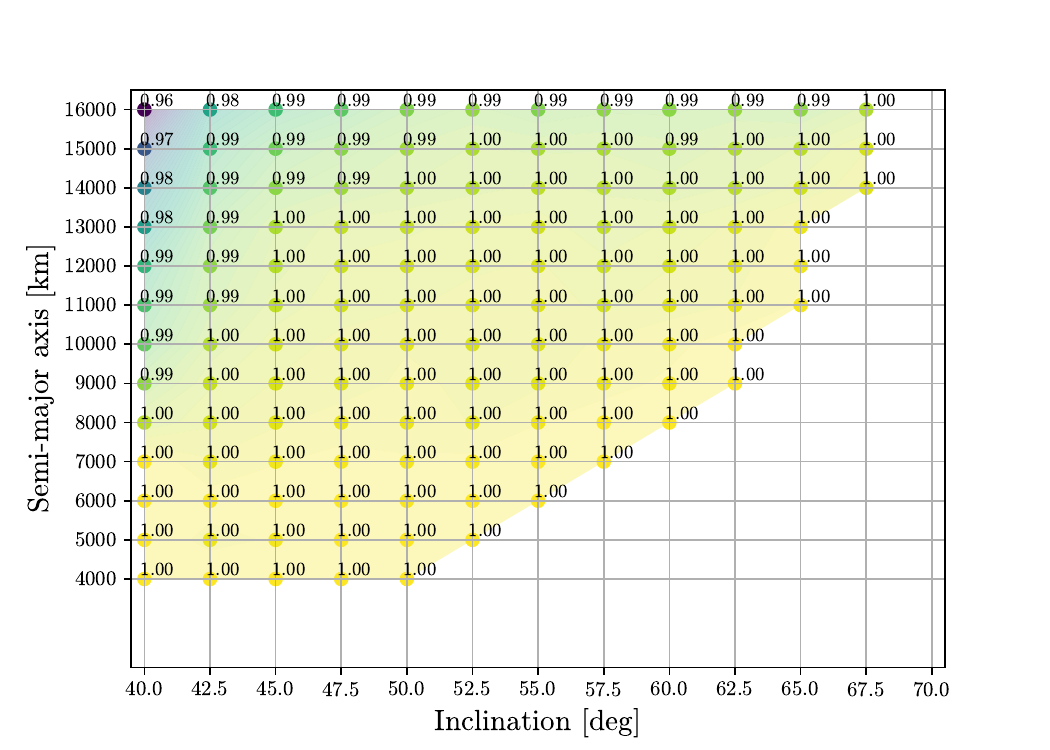}
    \subcaption{Ratio where $C/N_0 \geq $ 30 dB-Hz ($P_{tx}$ = 15 dBW)}
    \label{fig:cn0_Ptx15}
\end{minipage}
\hfill
\begin{minipage}
[b]{0.49\columnwidth}
    \centering
    \includegraphics[width=\textwidth]{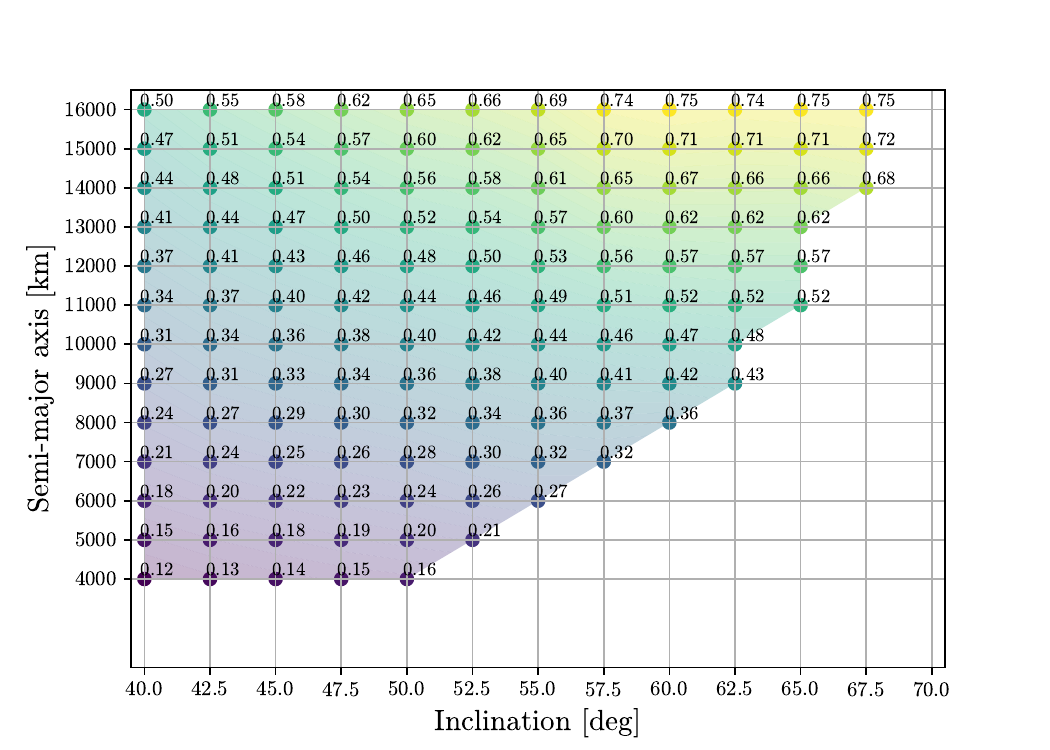}
    \subcaption{95th-percentile URE [m] ($P_{tx}$ = 15 dBW)}
    \label{fig:ure_Ptx15}
\end{minipage}
\caption{$C/N_0 \geq 30$ dB-Hz ratio and 95th-percentile URE for different transmit power levels (10 dBW and 15 dBW) and detection thresholds (30 dB-Hz).}
\label{fig:ure_cn0}
\end{figure}

Figure~\ref{fig:ure_cn0} shows (i) the fraction of samples (over all users and time) with \(C/N_0 \ge 30~\text{dB-Hz}\) and (ii) the corresponding 95th-percentile URE, for \(P_{\mathrm{tx}}\in\{8,15\}~\text{dBW}\).
With \(P_{\mathrm{tx}}=15~\text{dBW}\), the receiver maintains \(C/N_0 \ge 30~\text{dB-Hz}\) for most locations and times.
With \(P_{\mathrm{tx}}=8~\text{dBW}\), the fraction drops markedly at large semi-major axes and at both low and high inclinations.
At low inclinations (low \(e\)), satellites spend less time near apolune and the off-boresight angle to most users increases, reducing \(G_{\mathrm{tx}}\).
At high inclinations (high \(e\)), apolune occurs at high latitudes, so boresight alignment to much of the surface worsens over the orbit.

The 95th-percentile URE increases with \(a\) and \(i\), exceeding \(\sim 0.7~\text{m}\) at \(a=16{,}000~\text{km}\) and \(i=60^\circ\) for \(P_{\mathrm{tx}}=15~\text{dBW}\).
This trend reflects reduced received power at apolune in large, highly eccentric orbits.

\section{Insertion Cost Analysis}
\label{sec:insertion_cost}
In this section, we analyze the required lunar orbit insertion (LOI) $\Delta V$ for frozen orbits with different orbital elements.
Based on the work by~\citet{Ceresoli2025}, the required station-keeping $\Delta V$ for frozen orbits to maintain the semi-major axis is very small (below 1~m/s per year), so the LOI $\Delta V$ is the dominant contributor to the total $\Delta V$ budget.

\subsection{Trajectory Design Method}
We computed the LOI $\Delta V$ based on the method described in~\citet{ParkerAnderson2014}, as follows:
\begin{enumerate}
    \item Construct the target lunar orbit in the orbital-plane (OP) frame, with argument of periapsis $\omega = 90^{\circ}$, right ascension of ascending node (RAAN) $= 0^{\circ}$, and true anomaly $= 0^{\circ}$ (periapsis).
    \item Specify the epoch of LOI $t_{\mathrm{LOI}}$ and compute the satellite position and velocity in the Moon-centered inertial (MCI) frame.
    \item Specify the magnitude of the impulsive LOI maneuver $\Delta V_{\mathrm{LOI}}$. Apply the $\Delta V$ in the tangential direction to the LOI state.
    \item Propagate the state backward in time for 160~days. The force model includes the point-mass gravity of the Earth, Moon, and Sun. The ephemerides of the Earth, Moon, and Sun are obtained from JPL DE440.
    \item Identify the perigee and perilune passes in the trajectory:
    \begin{enumerate}
        \item If the trajectory flies by the Moon within 500~km, label it undesirable.
        \item The latest perigee passage that approaches within 500~km of Earth is considered the earliest opportunity to inject into that trajectory.
        \item If no low perigees are observed, the lowest perigee that approaches within 10{,}000~km of Earth (trajectories usable for real missions~\citep{ParkerAnderson2014}) is identified as the trans-lunar injection (TLI) location.
        \item Label the trajectory as undesirable if it does not approach within 10{,}000~km of Earth.
    \end{enumerate}
    \item Characterize the trajectory performance by recording:
    \begin{itemize}
        \item Characteristic energy ($C_3$) at TLI,
        \item Duration of the transfer,
        \item LOI $\Delta V$ magnitude.
    \end{itemize}
\end{enumerate}

Note that this method does not guarantee optimality in $\Delta V$ or duration. 
In practice, additional maneuvers at the TLI point or mid-course corrections may be required to insert into the desired lunar orbit, especially if multiple satellites are launched together and separated into different orbits.
However, this approach provides a systematic way to explore the trade-off space between $\Delta V$, duration, and $C_3$ for different lunar orbits.

\begin{figure}[ht!]
    \centering
    \includegraphics[width=0.4\linewidth]{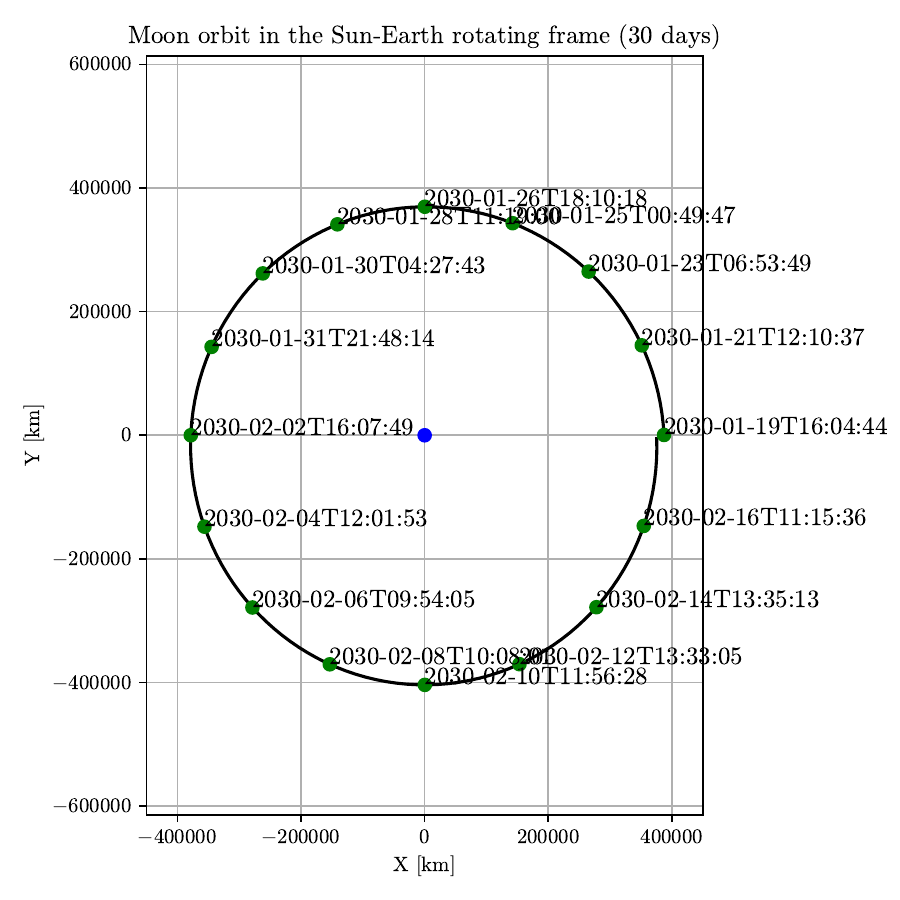}
    \caption{Sixteen insertion epochs (green dots) in the Earth-centered Sun-Earth fixed frame ($-x$ direction: Sun). The blue circle represents Earth; the gray circle, the Moon’s orbit.}
    \label{fig:dv_moon_phase}
\end{figure}

We swept 16 insertion epochs, equally spaced as the Moon orbits Earth (Figure~\ref{fig:dv_moon_phase}).
For LOI $\Delta V$, we swept over 951 values, sampled every 1~m/s from 50 to 1000~m/s. The obtained transfers are classified into three categories:
\begin{enumerate}
    \item \textbf{Direct Transfers:} These rely on two-body Earth–Moon dynamics. The spacecraft departs Earth on a hyperbolic trajectory directly intercepting the Moon. Such transfers typically require 4–6~days of flight time, depending on the Earth–Moon geometry.
    \item \textbf{Low-$C_3$ Transfers:} These use multiple Earth (and sometimes lunar) flybys to gradually raise apogee and reach the Moon. By gaining energy through flybys, the spacecraft achieves lunar transfer conditions with lower $C_3$ and smaller TLI $\Delta V$. Durations are longer than direct transfers but shorter than low-energy transfers.
    \item \textbf{Low-Energy Transfers:} Also called \textit{weak-stability boundary} transfers, these exploit the complex Earth–Moon–Sun dynamics (resonances and weak stability regions) to achieve capture with minimal $\Delta V$. The trade-off is a long transfer time (months) and higher $C_3$ ($\ge -1.0$~km$^2$/s$^2$) at TLI~\citep{ParkerAnderson2014}. Missions such as Hiten~\citep{Uesugi1996HITEN}, GRAIL~\citep{roncoli2010mission}, and SLIM~\citep{sakai2025moon} used such transfers.
\end{enumerate}

In this paper, transfers with duration $\le 7$~days are classified as direct, those with $C_3 \le -2.0$~km$^2$/s$^2$ as low-$C_3$, and those with maximum Earth distance $> 950{,}000$~km and $C_3 \ge -0.1$~km$^2$/s$^2$ as low-energy transfers.

\subsection{Results}
\begin{figure}[ht!]
    \centering
    \includegraphics[width=\linewidth]{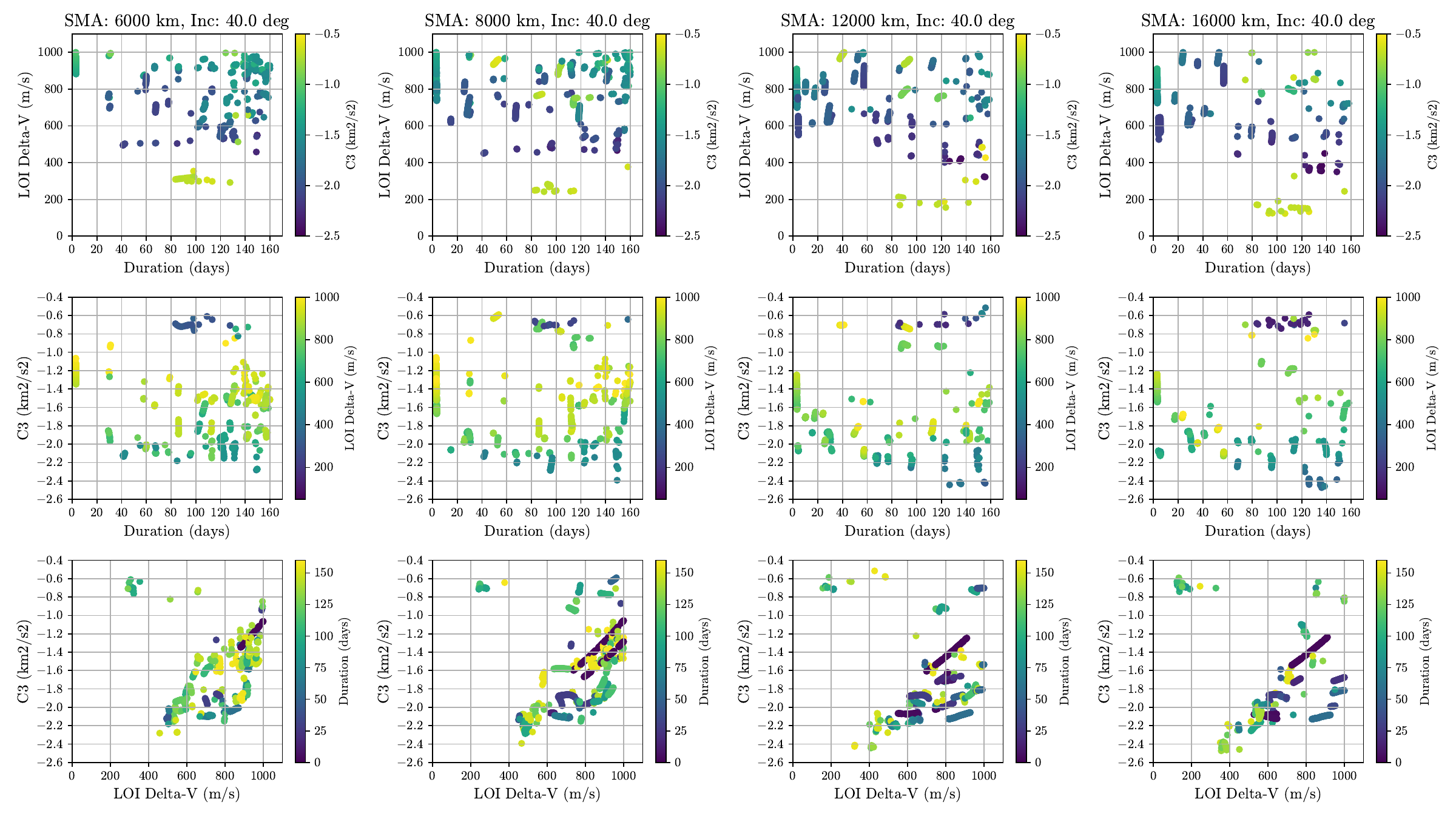}
    \caption{Trade-off between transfer duration, LOI $\Delta V$, and TLI $C_3$ for all feasible trajectories.}
    \label{fig:dv_tradeoff}
\end{figure}

The scatter plot of duration, LOI $\Delta V$, and TLI $C_3$ values for all feasible transfers at four semi-major axes (fixed $i = 40^{\circ}$) is shown in Figure~\ref{fig:dv_tradeoff}.
Direct transfers have $\Delta V$ ranging from 500–1000~m/s and $C_3$ from $-1.6$ to $-1.0$~km$^2$/s$^2$.
Low-$C_3$ transfers ($C_3 < -2.0$~km$^2$/s$^2$) require $\Delta V > 300$~m/s and durations typically $> 20$~days.
Low-energy transfers ($\Delta V \le 300$~m/s) require $> 80$~days and $C_3 > -0.8$~km$^2$/s$^2$.

\begin{figure}[ht!]
  \begin{minipage}[b]{\linewidth}
    \centering
    \includegraphics[width=\linewidth]{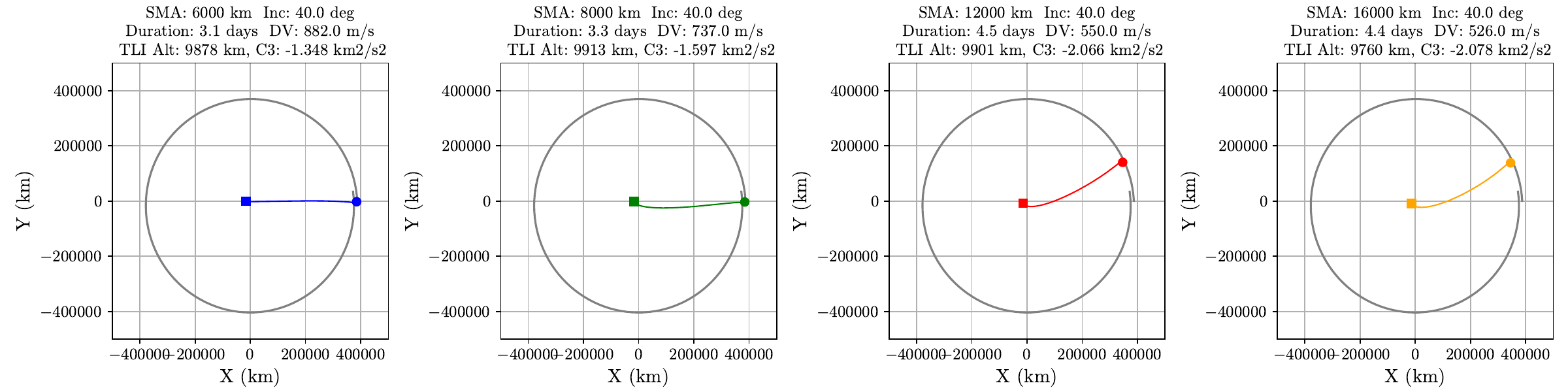}
    \subcaption{Minimum $\Delta V$ trajectories for direct transfers (TOF $\le 7$~days).}
  \end{minipage}

  \vspace{2mm}
  \begin{minipage}[b]{\linewidth}
    \centering
    \includegraphics[width=\linewidth]{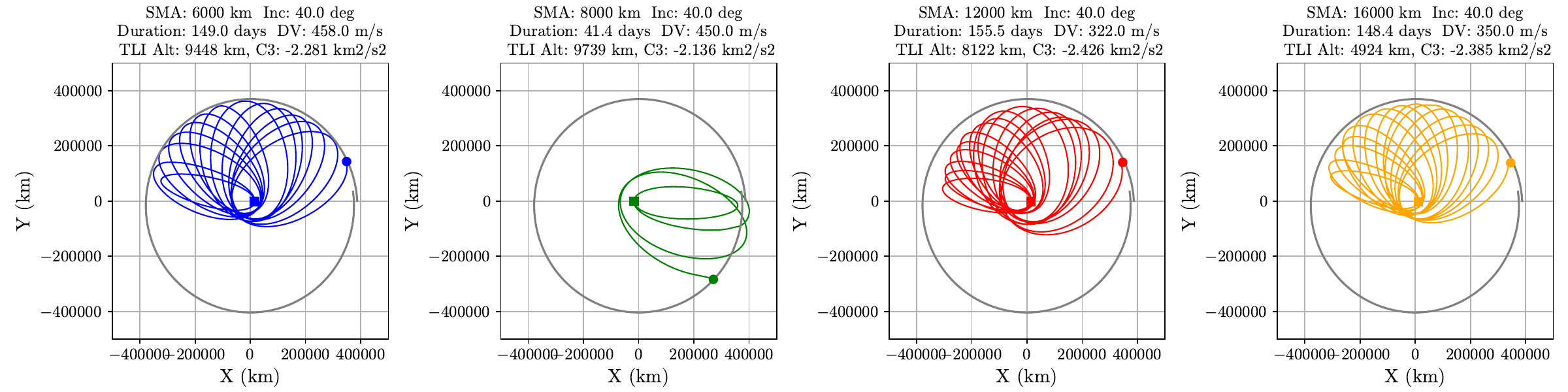}
    \subcaption{Minimum $\Delta V$ trajectories for low-$C_3$ transfers ($C_3 \le -2.0$~km$^2$/s$^2$).}
  \end{minipage}

  \vspace{2mm}
  \begin{minipage}[b]{\linewidth}
    \centering
    \includegraphics[width=\linewidth]{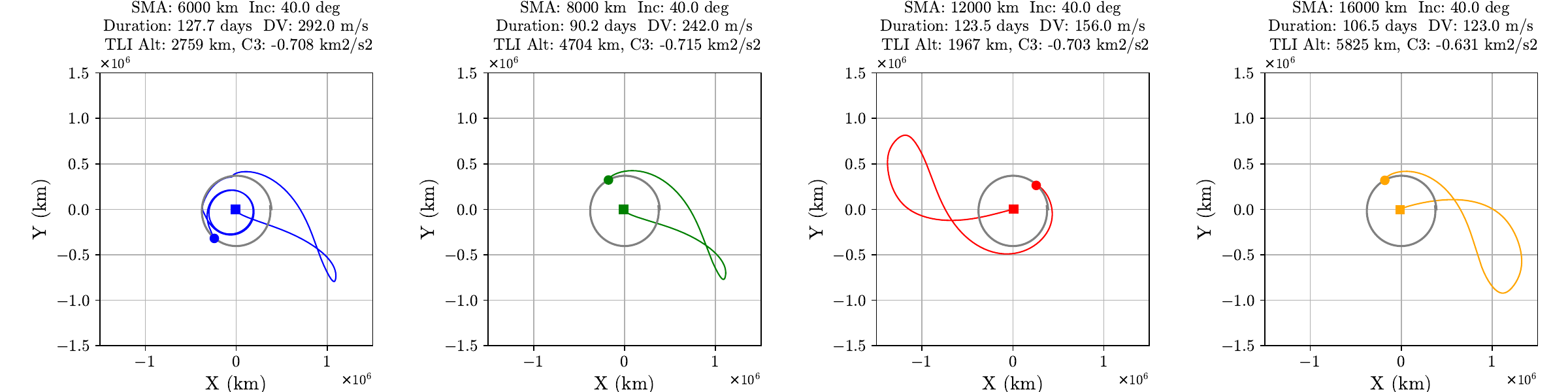}
    \subcaption{Minimum $\Delta V$ trajectories for low-energy transfers ($C_3 \ge -1.0$~km$^2$/s$^2$).}
  \end{minipage}

  \caption{Trajectories with minimum lunar orbit insertion $\Delta V$ for frozen orbits at four semi-major axes (6000–16{,}000~km). Inclination, RAAN, $\omega$, and mean anomaly at insertion are fixed to $0^{\circ}$, $0^{\circ}$, $90^{\circ}$, and $0^{\circ}$, respectively.
  Trajectories are shown in the Sun–Earth fixed frame ($-x$: Sun), with the Moon’s orbit in gray. Square markers indicate TLI points; circles indicate LOI points.}
  \label{fig:dv_trajectories}
\end{figure}

The trajectories with minimum LOI $\Delta V$ for frozen orbits with four semi-major axes (6000, 8000, 12{,}000, 16{,}000~km) are shown in Figure~\ref{fig:dv_trajectories}.
Direct transfers show simple Earth–Moon hyperbolic trajectories, low-$C_3$ transfers show multiple Earth flybys, and low-energy transfers follow complex paths far from Earth before lunar approach.

\begin{figure}[ht!]
  \begin{minipage}[b]{0.49\linewidth}
    \centering
    \includegraphics[width=\linewidth, trim={0mm 0mm 25mm 0mm}]{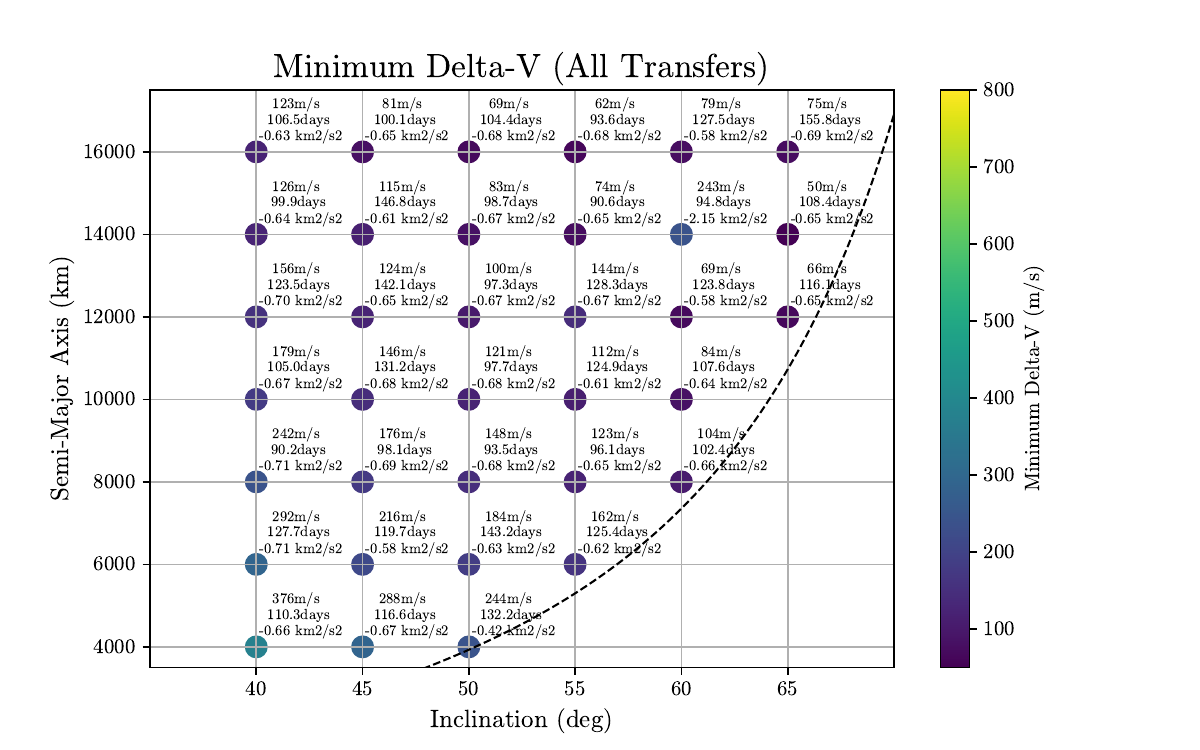}
    \subcaption{Minimum LOI $\Delta V$ for all transfers.}
  \end{minipage}
  \hfill
  \begin{minipage}[b]{0.49\linewidth}
    \centering
    \includegraphics[width=\linewidth, trim={0mm 0mm 25mm 0mm}]{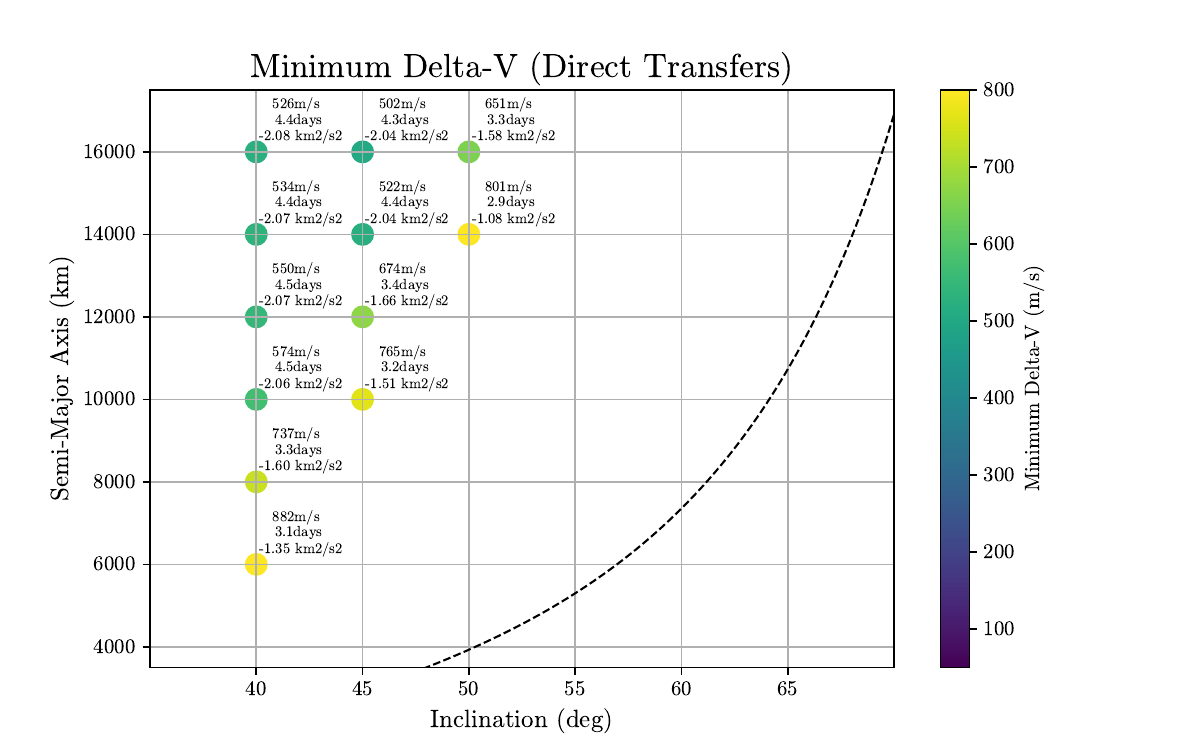}
    \subcaption{Minimum LOI $\Delta V$ for direct transfers.}
  \end{minipage}

  \begin{minipage}[b]{0.49\linewidth}
    \centering
    \includegraphics[width=\linewidth, trim={0mm 0mm 25mm 0mm}]{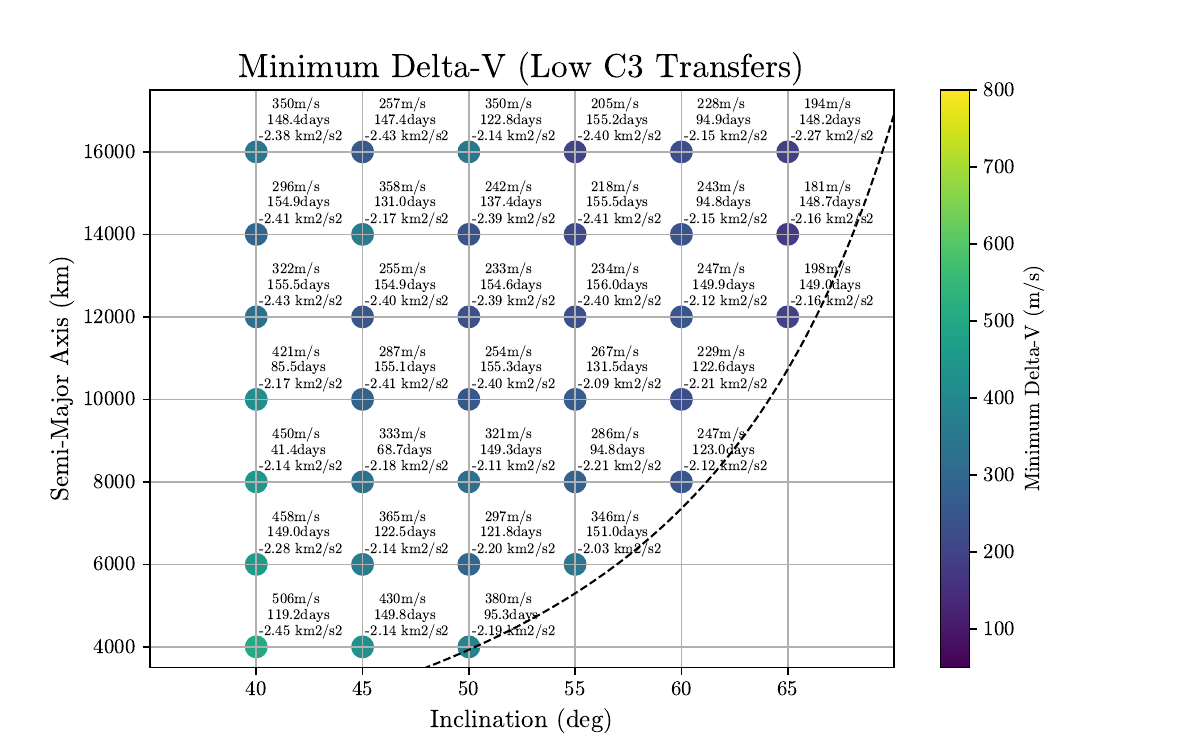}
    \subcaption{Minimum LOI $\Delta V$ for low-$C_3$ transfers.}
  \end{minipage}
  \hfill
  \begin{minipage}[b]{0.49\linewidth}
    \centering
    \includegraphics[width=\linewidth, trim={0mm 0mm 25mm 0mm}]{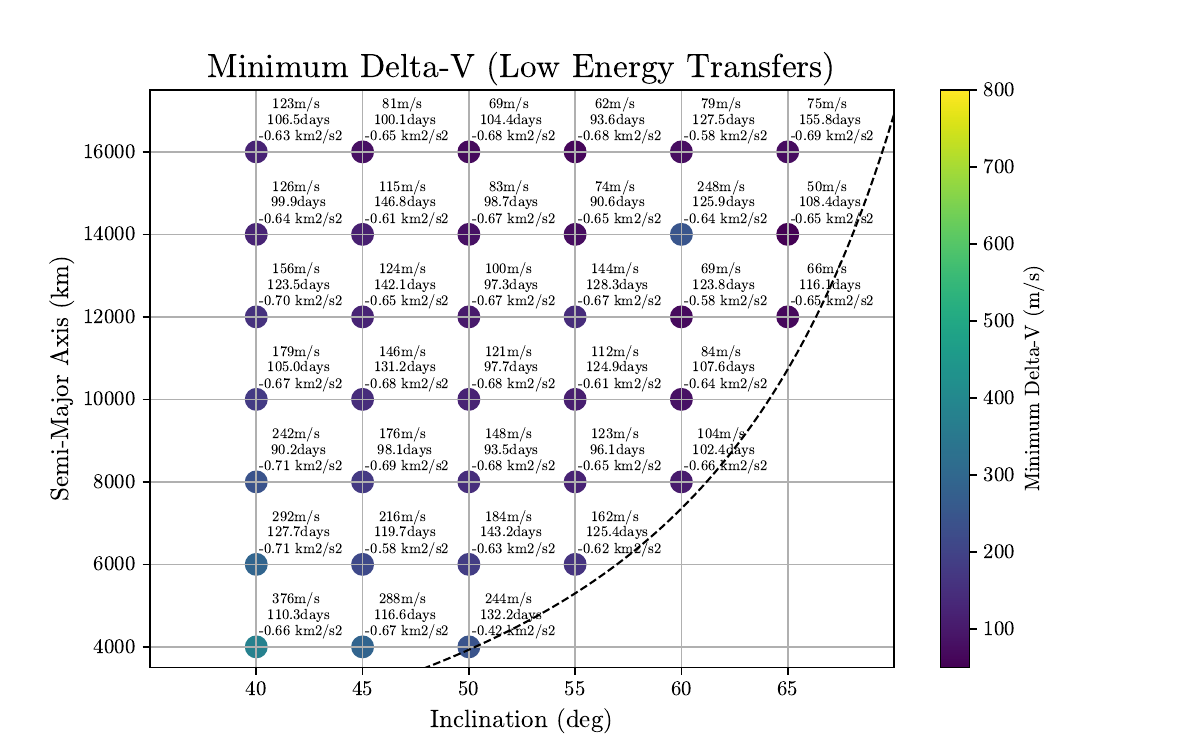}
    \subcaption{Minimum LOI $\Delta V$ for low-energy transfers.}
  \end{minipage}

  \caption{Minimum lunar orbit insertion $\Delta V$ across epochs for frozen orbits with varying semi-major axis and inclination.
  RAAN, $\omega$, and mean anomaly at insertion are fixed to $0^{\circ}$, $90^{\circ}$, and $0^{\circ}$, respectively.
  Text at each point indicates LOI $\Delta V$, time of flight, and TLI $C_3$.
  The dashed line marks the minimum semi-major axis (eccentricity) avoiding lunar impact.
  Blank points indicate no feasible transfer within the search space.}
  \label{fig:dv_transfers}
\end{figure}

The minimum LOI $\Delta V$ (over 16 epochs) for different semi-major axes and inclinations is shown in Figure~\ref{fig:dv_transfers}.
In general, $\Delta V$ decreases with larger semi-major axis, since the orbital energy difference is smaller.
No feasible direct transfers were found for inclinations above $50^{\circ}$ within the explored $\Delta V$ range.
Although inclination trends are less clear, higher inclinations tend to yield smaller $\Delta V$ since they bring the perilune closer to the Moon, enabling more efficient capture.

\section{Conclusion}
\label{sec:conclusion}
This paper presented a comprehensive trade-off analysis of lunar frozen orbit constellations for the Lunar Augmented Navigation Service (LANS). The study evaluated key performance metrics including coverage, position dilution of precision (PDOP), orbit determination error, receiver noise, and orbit insertion $\Delta V$, across a range of semi-major axes, inclinations, and right ascensions of the ascending node (RAAN).

The analysis revealed distinct coverage and PDOP characteristics among the three examined constellation types—S-ELFO Walker, NS-ELFO Walker, and CLFO Walker. Increasing the semi-major axis generally improved both polar and global coverage, though the optimal inclination depended on the constellation type and targeted coverage region. The S-ELFO Walker constellation provided superior performance in terms of polar coverage and PDOP, whereas the CLFO Walker constellation achieved the best global coverage and PDOP performance. The NS-ELFO Walker configuration exhibited weaker performance in both metrics but offered a practical balance for transitioning from pole-focused to global coverage constellations.

Orbit determination errors were found to decrease with smaller semi-major axes and inclinations.
For smaller semi-major axis orbits, RAAN played a notable role due to the varying geometry relative to Earth. 
Receiver noise exhibited similar dependencies, favoring lower altitudes and inclinations. 
In contrast, orbit insertion $\Delta V$ increased for configurations with lower semi-major axes and inclinations, highlighting a trade-off between coverage, orbit determination, URE, and insertion cost. Furthermore, the study compared different translunar injection (TLI) trajectories, showing that fast direct transfers demand higher $\Delta V$ but shorter flight times, whereas low-energy transfers substantially reduce $\Delta V$ at the expense of longer transfer durations.

The overall dependencies among performance metrics and orbital parameters are summarized in Table~\ref{tab:tradeoff_params}. These results emphasize the need to balance orbital geometry, insertion efficiency, and navigation performance in designing future lunar constellations. 

\begin{table}[htb]
\centering
\caption{Favorable Parameters for Different Performance Metrics}
\label{tab:tradeoff_params}
\begin{tblr}{
  colspec={X[c]X[c]X[c]X[c]},
  width=\textwidth,
  row{even} = {white, font=\small},
  row{odd} = {bg=black!10, font=\small},
  row{1} = {bg=black!20, font=\bfseries\small},
  hline{Z} = {1pt, solid, black!60},
  rowsep=3pt
}
\textbf{Metric} & \textbf{Semi-major Axis} & \textbf{Inclination (Eccentricity)} & \textbf{RAAN} \\
Pole Coverage & Larger & Larger & -- \\
Global Coverage & Larger & Smaller & -- \\
DOP & Larger & Medium (45$^{\circ}$--55$^{\circ}$) & -- \\
OD Error & Smaller & Smaller & Depends on Earth Direction \\
Receiver Noise & Smaller & Smaller & -- \\
Orbit Insertion $\Delta V$ & Larger & Larger & Depends on epoch \\
\end{tblr}
\end{table}

Future work will extend this analysis by incorporating different orbit determination and time synchronization (ODTS) methods—including terrestrial sidelobe GNSS and inter-satellite link-based approaches—and by integrating spacecraft subsystem modeling for mass and power budgeting. Additionally, multi-objective optimization will be applied to refine constellation architectures and staged deployment strategies. Collectively, this work provides a foundational framework for designing robust and efficient lunar navigation constellations that optimize coverage, positioning accuracy, and operational cost.

\section*{acknowledgements}
This material is based upon work supported by The Nakajima Foundation.

\nocite{*}
\printbibliography[title=References]

\end{document}